\documentclass[nologo,11pt,a4paper]{ETHpaper}

\usepackage{graphicx, epsfig, amsmath, amssymb}
\usepackage[square,numbers,sort&compress]{natbib}
\usepackage{latexsym}
\usepackage{subfigure}
\usepackage{pstricks,pst-node,pst-plot}

\begin{document}

\title{Nonlinear Voter Models: \\ The Transition from Invasion to
  Coexistence}
\titlealternative{Nonlinear Voter Models: The Transition from Invasion to
  Coexistence}

\author{Frank Schweitzer$^\star$\footnote{Corresponding author:
    \url{fschweitzer@ethz.ch}}, Laxmidhar Behera$^\dagger$}

\authoralternative{Frank Schweitzer, Laxmidhar Behera}

 \address{$^\star$Chair of Systems Design, ETH  Zurich, Kreuzplatz 5,
   8032 Zurich, Switzerland \\
         $^\dagger$Department of Electrical Engineering,
    Indian Institute of Technology, Kanpur 208 016, India}

  \reference{\emph{European Physical Journal B}, vol \textbf{67} (2009). 
}

\www{\url{http://www.sg.ethz.ch}}

\makeframing
\maketitle

\begin{abstract}
In nonlinear voter models the transitions between two states
  depend in a nonlinear manner on the frequencies of these states in the
  neighborhood. We investigate the role of these nonlinearities on the
  global outcome of the dynamics for a homogeneous network where each
  node is connected to $m=4$ neighbors. The paper unfolds in two
  directions. We first develop a general stochastic framework for
  frequency dependent processes from which we derive the macroscopic
  dynamics for key variables, such as global frequencies and
  correlations. Explicit expressions for both the mean-field limit and
  the pair approximation are obtained. We then apply these equations to
  determine a phase diagram in the parameter space that distinguishes
  between different dynamic regimes.  The pair approximation allows us to
  identify three regimes for nonlinear voter models: (i) complete
  invasion, (ii) random coexistence, and -- most interestingly -- (iii)
  correlated coexistence. These findings are contrasted with predictions
  from the mean-field phase diagram and are confirmed by extensive
  computer simulations of the microscopic dynamics. 

\end{abstract}

\date{\today}

\newcommand{\mean}[1]{\left\langle #1 \right\rangle}
\newcommand{\abs}[1]{\left| #1 \right|}
\newcommand{\eqn}[1]{eqn. (\ref{#1})}
\newcommand{\Eqn}[1]{Eqn. (\ref{#1})}
\newcommand{\sect}[1]{Sect. \ref{#1}}
\newcommand{\eqs}[2]{eqs. (\ref{#1}), (\ref{#2})}
\newcommand{\pic}[1]{Fig. \ref{#1}}
\newcommand{\name}[1]{{\rm #1}}
\newcommand{\bib}[4]{\bibitem{#1} {\rm #2} (#4): #3.}
\newcommand{\ul}[1]{\underline{#1}}
\renewcommand{\epsilon}{\varepsilon}
\newcommand{\eps}{\varepsilon}
\renewcommand*{\=}{{\kern0.1em=\kern0.1em}}
\renewcommand*{\-}{{\kern0.1em-\kern0.1em}}
\newcommand*{\+}{{\kern0.1em+\kern0.1em}}
\frenchspacing
\setlength{\fboxsep}{1pt}
\renewcommand{\textfraction}{0.02}
\renewcommand{\topfraction}{0.98}
\renewcommand{\bottomfraction}{0.98}
\renewcommand{\floatpagefraction}{0.98}

\section{Introduction}
\label{1}

In biological systems, the survival of a species depends on the
frequencies of its kin and its foes in the environment
\citep{antonovics-kareiva-88,molofsky99}. In some cases, the chance of
survival of a certain species \emph{improves} as the frequency of its
kind increases, since this might enhance the chance for reproduction or
other benefits from group interaction. This is denoted as \emph{positive}
frequency dependence. In other cases a \emph{negative} frequency
dependence, that is the increase of the survival chance with
\emph{decreasing} frequency, is observed. This is the case, when
individuals compete for rare ressources. Moreover, negative frequency
dependence is known to be important for maintaining the genetic diversity
in natural populations \citep{rohlf-schnell-71,kimura-weiss-64}.

Frequency dependent dynamics are not only found in biological systems,
but also in social and economic systems \citep{albin-75, schelling-69,
  weidl-94-b, fs-holyst-00, nowak-et-00, holyst-kacp-fs-00,Krapivsky:03,
 castellano2007rmp}.
In democracies, a simple example is a public vote, where the winning
chances of a party increase with the number of supporters
\citep{costa-99,bernardes:2002}.  In economics, e.g.  the acceptance of a
new products may increase with the number of its users \citep{oomes-02}.
In stock markets, on the other hand, positive and negative frequency
dependencies may interfere. For instance, the desire to buy a certain
stock may increase with the orders observed from others, a phenomenon
known as the \emph{herding effect}, but it also may decrease, because
traders fear speculative bubbles.

In general, many biological and socio-economic processes are governed by
the frequency dependent adoption of a certain behavior or strategy, or
simply by frequency dependent reproduction. In order to model such
dynamics more rigorously (but less concrete), different versions of
\emph{voter models} have been investigated.  The voter model denotes a
simple binary system comprised of $N$ {\em voters}, each of which can be
in one of two states (where \emph{state} could stand for opinion,
attitude, or occupation etc.), $\theta_{i}=\{0,1\}$. Here, the transition
rate $w(\theta|\theta^{\prime})$ from state $\theta^{\prime}$ to state
$\theta$ is assumed to be proportional to the frequency $f_{\theta}$. In
this paper, we extend this approach by assuming a \emph{nonlinear} voter
model, where the frequency dependence of the transition rate,
$w(\theta|\theta^{\prime})=\kappa(f)\,f_{\theta}$, includes an additional
nonlinearity expressed in terms of the (frequency dependent) prefactor
$\kappa$.

Linear voter models have been discussed for a long time in mathematics
\citep{cg86b, holley-liggett-75, lig94, liggett-99}. Recently, they
gained more attention in statistical physics \citep{redner2001,
  dallasta2007est, castellano2005, sood2005, maxi2, suchecki2005,
  castellano2003, dornic2001, moore-97, vazquez08, stark08,
  Krapivsky:03,bennaim2003bap,PhysRevE.53.3078} because of some
remarkable features in their dynamics described in
Sect. \ref{sec:vm}. But voter models also found the interest of
population biologists \citep{molofsky99,
  keitt01:_allee_effec_invas_pinnin_border,
  kendall00:_disper_envir_correl_spatial_synch_popul_dynam,
  neuhauser99:_ances_graph_gene_geneal_frequen_depen_selec, Durrett:94,
  pacala85:_neigh_model}

Dependent on how the frequency $f_{\theta}$ is estimated, one can
distinguish global from local voter models. In the latter case the
transition is governed only by the local frequency of a certain state in
a given neighborhood. In contrast to global (or mean-field) models, this
leads to local effects in the dynamics, which are of particular interest
in the current paper.  If space is represented by a two-dimenional
lattice and each site is occupied by just one individual, then each
species occupies an amount of space proportional to its presence in the
total population. Local effects such as the occupation of a neighborhood
by a particular species or the adoption of a given opinion in a certain
surrounding, can then be observed graphically in terms of domain
formation.  This way, the invasion of species (or opinions) in the
environment displays obvious analogies to spatial pattern formation in
physical systems.

Physicists have developed different spatial models for such processes.
One recent example is the so-called ``Sznajd model''
\citep{bernardes:2002,lb-fs-03,slanina2003} which is a simple cellular
automata (CA) approach to consensus formation (i.e. complete invasion) among
two opposite opinions (described by spin up or down).  In
\citep{lb-fs-03}, we have shown that the Sznajd model can be completely
reformulated in terms of a linear voter model, where the transition rates
towards a given opinion are directly proportional to the frequency of the
respective opinion of the \emph{second-nearest} neighbors and independent
of the nearest neighbors.

Other spatial models are proposed for game-theoretical interactions among
nearest neighbors \citep{szabo-et-00, Nakamaru:97}. Here, the dynamics
are driven by local payoff differences of adjacent players, which
basically determine the nonlinearity $\kappa(f)$.  Dependent on these
payoff differences, we could derive a phase diagram with five regimes,
each characterized by a distinct spatio-temporal dynamic
\citep{fs-lb-acs-02}. The corresponding spatial patterns range from
complete invasion to coexistence with large domains, coexistence with
small clusters, and spatial chaos.

In this paper, we are interested in the local effects of frequency
dependent dynamics in a homogeneous network, where each site has $m=4$
nearest neighbors. In this case, the nonlinearity $\kappa(f)$ can be
simply expressed by two constants, $\alpha_{1}$, $\alpha_{2}$. This is a
special form of a nonlinear voter model, which for
$\alpha_{1}<\alpha_{2}<0.5$ also includes majority voting and for
$\alpha_{1}>\alpha_{2}>0.5$ minority voting. We investigate the dynamics
of this model both analytically and by means of computer simulations on a
two-dimensional stochastic CA (which is a special form of a homogeneous
network with $m=4$). The latter one was already studied in
\citep{molofsky99}, in particular there was a phase diagram obtained via
numerical simulations. In our paper, we go beyond that approach by
deriving the phase diagram from an analytical approximation, which is
then compared with our own simulations.

In Sects. \ref{sec:formal}, \ref{sec:micro} we introduce the microscopic
model of frequency dependent invasion and demonstrate in
Sects. \ref{sec:a1a2}, \ref{2.4} the role of $\alpha_{1}$, $\alpha_{2}$
by means of characteristic pattern formation. Based on the microscopic
description, in Sect. \ref{3.1} we derive the dynamics for the global
frequency $x(t)$, which is a macroscopic key variable. An analytical
investigation of these dynamics is made possible by pair approximation,
Sect. \ref{3.2}, which results in a closed-form description for $x(t)$
and the spatial correlations $c_{1|1}(t)$. In Sect. \ref{4.1}, we verify
the performance of our analytical approximations by comparing them with
averaged CA computer simulations.  The outcome of the comparison allows
us to derive in Sect. \ref{4.3} a phase diagram in the
$(\alpha_{1},\alpha_{2})$ parameter space, to distinguish between two
possible dynamic scenarious: (i) complete invasion of one of the species,
with formation of domains at intermediate time scales, and (ii) random
spatial coexistence of two species.  A third dynamic regime, the
nonstationary coexistence of the two species on long time scales together
with the formation of spatial domains, can be found in a small, but
extended region that separates the two dynamic regimes mentioned
above. We further discuss in Sect. \ref{5} that the usual distinctions
for the dynamics, such as positive or negative frequency dependence, do
not necessarily coincide with the different dynamic regimes.  Instead,
for positive frequency dependence, all of the three different dynamic
regimes (and the related spatio-temporal patterns) are observed. In the
Appendix, calculation details for the pair approximation are given.

\section{Formal Approach to Voter Models}
\label{sec:formal}

\subsection{Defining the system}
\label{2.1}

We consider a model of two species labeled by the index $\sigma=\{0,1\}$.
The total number of individuals is constant, so the global frequency
$x_{\sigma}$ (or the share of each species in the total population) is
defined as:
\begin{eqnarray}
  \label{nconst}
  N&=&\sum_{\sigma}N_{\sigma}=N_{0}+N_{1}= \mathrm{const.} \nonumber \\
 x_{\sigma}&=&\frac{N_{\sigma}}{N} \;; \quad x \equiv x_{1} = 1-x_{0}
\end{eqnarray}
In the following, the variable $x$ shall refer to the global frequency of
species 1.

The individuals of the two species are identified by the index $i \in N$
and can be seen as nodes of a network.  A discrete value
$\theta_{i}\in\{0,1\}$ indicates whether the node is occupied by species
0 or 1. The network topology (specified later) then defines the nearest
neighbors $i_{j}$ of node $i$. In this paper, we assume homogeneous
networks where all nodes have the same number $m$ of nearest neighbors.
For further use, we define the local occupation $\ul{\theta}_{i}$ of the
nearest neighborhood (without node $i$) as:
\begin{equation}
  \label{occupat}
\ul{\theta}_{i}= \{\theta_{i_1},\theta_{i_2},...,\theta_{i_{m}}\}  
\end{equation}
A specific realization of this distribution shall be denoted as
$\ul{\sigma}$, while the function $\ul{\eta}_{i}(\ul{\sigma})$ assigns
$\ul{\sigma}$ to a particular neighborhood $\ul{\theta}_{i}$:
\begin{equation}
  \begin{aligned}
 \ul{\sigma}=&\{\sigma_{1},\sigma_{2},...,\sigma_{m}\}    \\
\ul{\eta}_{i}(\ul{\sigma}) =& \{\theta_{i_1}\=\sigma_{1},
\theta_{i_2}\=\sigma_{2},...,\theta_{i_{m}}\=\sigma_{m}\}    
  \end{aligned}
  \label{sigm01}
\end{equation}
For later use, it is convenient to define these distributions also for
the nearest neighborhood \emph{including} node $i$:
\begin{equation}
  \label{occupat0}
  \begin{aligned}
\ul{\theta}_{i}^{0}=&
\{\theta_{i},\theta_{i_1},\theta_{i_2},...,\theta_{i_{m}}\} = 
\ul{\theta}_{i} \cup \{\theta_{i}\} \\
   \ul{\sigma}^{0}=&\{\sigma,\sigma_{1},\sigma_{2},...,\sigma_{m}\} \\
\ul{\eta}^{0}_{i}(\ul{\sigma}^{0}) =&
\{\theta_{i}\=\sigma,\theta_{i_1}\=\sigma_{1},
\theta_{i_2}\=\sigma_{2},...,\theta_{i_{m}}\=\sigma_{m}\}    
  \end{aligned}
\end{equation}
For $m=4$, $\ul{\sigma}^{0}$ denotes a binary string, e.g. $\{01001\}$,
where the first value $\sigma$ refers to the center node and the other
values $\sigma_{j}\in\{0,1\}$ indicate the particular values of the
nearest neighbors. The assignment of these values to a particular
neighborhood $\ul{\theta}_{i}^{0}$ of node $i$ is then described by
$\ul{\eta}^{0}_{i}(\ul{\sigma}^{0})$.

In the voter model described in the following section, the dynamics of
$\theta_{i}$ is governed by the \emph{occupation distribution} of the
\emph{local neighborhood}. that surrounds each node $i$. Using a
stochastic approach, the probability $p_i(\theta_i,t)$ to find node $i$
in state $\theta_{i}$ therefore depends in general on the
local occupation distribution $\ul{\theta}_{i}$ of the neigborhood
(\eqn{occupat}, in the following manner:
\begin{equation}
\label{eq:marg_dist}
p_i(\theta_i,t)=\sum_{\ul{\theta}_{i}^{\prime}}
p(\theta_i,\ul{\theta}_{i}^{\prime}, t)
\end{equation}
Hence, $p_i(\theta_i,t)$ is defined as the marginal distribution of
$p(\theta_i,\ul{\theta}_{i}, t)$, where $\ul{\theta}_{i}^{\prime}$ in
\eqn{eq:marg_dist} indicates the summation over all possible realizations
of the local occupation distribution $\ul{\theta}_{i}$, namely $2^{m}$
different possibilities.

For the time dependent change of $p_i(\theta_i,t)$ we assume the
following master equation: 
\begin{eqnarray}
\label{master}
\frac{d}{dt}p_i(\theta_i,t)&=&\sum_{\ul{\theta}_{i}^{\prime}} \Big[
w(\theta_i|(1\-\theta_i),\ul{\theta}_{i}^{\prime})\;
p(1\-\theta_i,\ul{\theta}_{i}^{\prime},t) \nonumber \\
& & \qquad - w(1\-\theta_i|\theta_i,\ul{\theta}_{i}^{\prime})\;
p(\theta_i,\ul{\theta}_{i}^{\prime},t)\Big]
\end{eqnarray}
where $w(\theta_i|(1\-\theta_i),\ul{\theta}_{i})$ denotes the transition
rate for state $(1\-\theta_i)$ of node $i$ into state $\theta_{i}$ in the
next time step under the condition that the local occupation distribution
is given by $\ul{\theta}_{i}$. The transition rate for the reverse
process is $w(1\-\theta_i|\theta_i,\ul{\theta}_{i})$. Again, the
summation is over all possible realizations of $\ul{\theta}_{i}$, denoted
by $\ul{\theta}_{i}^{\prime}$. It remains to specify the transition
rates, which is done in the following section.

\subsection{Linear and Nonlinear Voter Models}
\label{sec:vm}

Our dynamic assumptions for the change of an individual state
$\theta_{i}$ are taken from the so-called voter model (see also Sect.
\ref{1}), abbreviated as VM in the following. The dynamics is given by
the following update rule: A voter, i.e. a node $i\in N$ of the network,
is selected at random and adopts the state of a randomly chosen nearest
neighbor ${j}$. After $N$ such update events, time is increased by 1.

The probability to choose a voter with a given state $\sigma$ from the
neighborhood $i_{j}$ of voter $i$ is directly proportional to the
relative number (or frequency) of voters with that particular state in
that neigborhood. Let us define the \emph{local frequencies} in the
neighborhood as:
\begin{equation}
  \label{sum}
f_{i}^{\sigma} = \frac{1}{m}\sum_{j=1}^{m} \delta_{\sigma\theta_{i_{j}}} \;;\quad
f_{i}^{(1\-\sigma)} = 1 - f_{i}^{\sigma}
\end{equation}
where $\delta_{xy}$ is the Kronecker delta, which is 1 only for $x=y$ and
zero otherwise. Then the transition rate of voter $i$ to change its state
$\theta_{i}$ does not explicitly depend on the local distribution
$\ul{\theta}_{i}$, but only on the \emph{occupation frequency}
$f_{i}^{\sigma}$, i.e. on the number of nodes occupied by either 0 or 1
in the neighborhood of size $m$.  Hence, the VM describes a frequency
dependent dynamics: the larger the frequency of a given state in the
neighborhood, the larger the probability of a voter to switch to that
particular state \emph{if} it is not already in that state. I.e. the
transition rate $w(1\-\theta_{i}|\theta_{i}\=\sigma,f_{i}^{\sigma})$, to
\emph{change} state $\theta$ increases only with the local frequency of
\emph{opposite} states, $f_{i}^{1-\sigma}$, in the neighborhood:
\begin{equation}
  \label{eq:linvm}
  w(1\-\theta_{i}|\theta_{i}\=\sigma,f_{i}^{\sigma})= \gamma f_{i}^{1-\sigma}
\end{equation}
The prefactor $\gamma$ determines the time scale of the transitions and is
set to $\gamma=1$.  Eqn. (\ref{eq:linvm}, describes the dynamics of the
\emph{linear} VM because, according to the above update rule, the rate to
\emph{change} the state is directly proportional to the frequency.

The linear (or standard) VM has two remarkable features. First, it is
known that, starting from a random distribution of states, the system
always reaches a completely ordered state, which is often referred to as
\emph{consensus} in a social context, or complete \emph{invasion} in a
population biology context.  As there are individuals with two different
states, the complete ordered state can be either all 0 or all 1. Which of
these two possible attractors of the dynamics is eventually reached,
depends (in addition to stochastic fluctuations) on the initial global
frequency, i.e. $x(t=0)$. It has been shown that, for an ensemble
average, the frequency of the outcome of a particular consensus state $1$
is equal to the initial frequency $x(t=0)$ of state $1$. This second
remarkable feature is often denoted as conservation of magnetization,
where the magnetization is defined as $M(t)=x_{1}(t)-x_{0}(t)=2x(t)-1$.
Hence, consensus means $\abs{M}=1$. Thus we have the interesting
situation that, for a single realization, the dynamics of the linear VM
is a fluctuation driven process that, for finite system sizes, always
reaches consensus, whereas on average the outcome of the consensus state
is distributed as $x(0)$. 

The (only) interesting question for the linear VM is then how long it may
take the system to reach the consensus state, dependent on the system
size $N$ and the network topology.  The time to to reach consensus,
$T_{\kappa}$, is obtained through an average over many realizations.  As
the investigation of $T_{\kappa}$ is not the focus of our paper (see
\citep{frachenbourg1996, krapivsky1992kmm, maxi2}), we just mention some
known results for the linear VM : One finds for one-dimensional regular
lattices ($d=1$) $T_\kappa\propto N^2$ and for two-dimensional regular
lattices ($d=2$) $T_\kappa\propto N \log N$. For $d>2$ the system does
not always reach an ordered state in the thermodynamic limit. In finite
systems, however, one finds $T_{\kappa} \sim N$.

While the linear VM has some nice theoretical properties, it also has
several conceptual disadvantages when applying the model to a social or
population biological context. First of all, the ``voters'' do not vote
in this model, they are subject to a random (but frequency based)
assignment of an ``opinion'', without any choice. Secondly, the state of
the voter under consideration does not play any role in the dynamics.
This can be interpreted in a social context as a (blind) herding
dynamics, where the individuals just adopt the opinion of the majority.
In a population model of two competing species, it means that individuals
from a minority species may be replaced by those from a majority species
without any resistance.

In order to give voter $i$ at least some weight compared to the influence
of its neighbors $i_{j}$, one can simply count its state $\theta_{i}$
into the local frequency $f_{i}^{\sigma}$, i.e. instead of eqs.
(\ref{occupat}), (\ref{sigm01}) we may use eqn. (\ref{occupat0}). Using
for voter $i$ the notation $\theta_{i}\equiv \theta_{i_{0}}$ (i.e
$j\=0$), we can still use eqn. (\ref{eq:linvm}) for the transition rates,
with the noticable difference that the local frequency $f_{i}^{\sigma}$
of eqn.  (\ref{sum}) is now calculated from a summation that starts with
$j=0$.  The explicit consideration of $\theta_{i}$ thus has the effect
of adding some \emph{inertia} to the dynamics. In fact, extending the
summation to $j=0$ multiplies the transition rate, eqn. (\ref{eq:linvm}),
by a factor $m/(m+1)$, where $m$ is the number of nearest neighbors.
I.e., for $m=4$ a local configuration $\ul{\sigma}^{0}=\{01001\}$ would
lead to a transition rate $w(1|0)=0.5$ \emph{without} the additional
inertia, but $w(1|0)=0.4$ by counting in the state of voter $i$. I.e.,
taking into account the state of voter $i$ considerably reduces the
transition rate towards the opposite state. 

We find it useful for conceptual reasons to include some resistance into
the model and therefore will use from now on the description which takes
the current state of voter $i$ into account. This also has the nice
advantage that for the case $m=4$, which describes e.g. square lattices,
we avoid stalemate situations, $w(1-\theta|\theta)=0.5$. However, we note
that the addition of the constant resistance does not change the
dynamics of the model, as it only adjusts the \emph{time scale} towards a
new factor $\gamma^{\prime}=(m/m+1)\,\gamma$. So, keeping $m$ constant
and equal for all voters, we can rescale $\gamma^{\prime}=1$. 

We note that there are of course other ways to give some weight to the
opinion of voter $i$. In \citep{stark08,stark08b}, we have discussed a
modified VM, where voters additionally have an inertia $\nu_i \in [0,1]$
which leads to a decrease of the transition rate to change their state:
\begin{equation}
w^{R}(1-\theta_{i}|\theta_{i})=(1-\nu_{i})\; w(1-\theta_{i}|\theta_{i})
\label{eq:reluc}
\end{equation}
Here $w(1-\theta|\theta)$ is given by the linear VM, eqn.
(\ref{eq:linvm}). The individual inertia $\nu_{i}$ is evolving over time
by assuming that it increases with the persistence time $\tau_i$ the
voter has been keeping its current state.  While this intertia
may slow down the microscopic dynamics of the VM and thus may increase
the time to reach consensus, $T_{\kappa}$, we found the counterintuitive
result that under certain circumstances a decelerated microdynamics may
even accelerate the macrodynamics of the VM, thus decreasing $T_{\kappa}$
compared to the linear VM.

\begin{figure}[htbp]
  \centering
  \includegraphics[width=7cm]{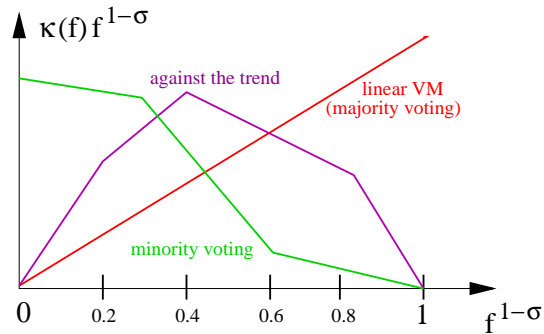}
  \caption{Different nonlinear frequency dependencies for eqn.
    (\ref{eq:nonlinvm}). Note the piecewise linear functions, as the
    number of neighbors $m$ and thus the frequencies $f$ have mostly
    discrete values.}
  \label{fig:nonlin}
\end{figure}

The addition of a nonlinear inertia to the VM, eqn. (\ref{eq:reluc}), is
a special case for turning the linear VM into a nonlinear one (wheras the
fixed resistence would not change the linear VM). In general, nonlinear
VM can be expressed as 
\begin{equation}
  \label{eq:nonlinvm}
  w(1\-\theta_{i}|\theta_{i}\=\sigma,f_{i}^{\sigma})= \kappa(f)\; f_{i}^{1-\sigma}
\end{equation}
where $\kappa(f)$ is a nonlinear, frequency dependent function describing
how voter $i$ reacts on the occurence of opposite ``opinions'' in its
immediate neighborhood. Fig. \ref{fig:nonlin} shows some possible
examples which have their specific meaning in a social context. Whereas
any function $\kappa(f)=\mathrm{const.}>0$ describes the linear VM, i.e.
a majority voting or herding effect, a decreasing $\kappa(f)$ means
minority voting, i.e. the voter tends to adopt the opinion of the
minority. Nonmonotonous $\kappa(f)$ can account for voting against the
trend, i.e. the voter adopts an opinion as long as this is not already
the ruling opinion -- a phenomenon which is important e.g. in modeling
the adoption of fashion.  An interpretation of these functions in a
population biology context will be given in Sect. \ref{sec:a1a2}

In conclusion, introducing the nonlinear response function $\kappa(f)$
will allow us to change the global dynamics of the linear VM. Instead of
reaching always consensus, i.e. the exclusive domination of one
``opinion'' or species, we may be able to observe some more interesting
macroscopic dynamics, for example the coexistence of both states. It is
one of the aims of this paper to find out, under which specifications of
$\kappa(f)$ we may in fact obtain a dynamic transition that leads to a a
structured, but not fully ordered state instead of a completely ordered
state.

\section{Stochastic Dynamics of the Voter Model}
\label{2.2}

\subsection{Microscopic Dynamics}
\label{sec:micro}

In order to give a complete picture of the dynamics of the nonlinear VM,
we have to derive the stochastic dynamics for the \emph{whole} system of
$N$ nodes, whereas eqn. (\ref{master}) gives us ``only'' the \emph{local}
dynamics in the vicinity of a particular voter $i$. For $N$ voters, the
distribution of states is given by
\begin{equation}
  \label{vector}
  {\Theta}=\{\theta_1,\theta_2,...,\theta_N\}
\end{equation}
Note that the state space $\Omega$ of all possible configurations is of
the order $2^{N}$.  In a stochastic model, we consider the probability
$p({\Theta},t)$ of finding a particular configuration at time $t$.
If $t$ is measured in discrete time steps (generations) and the network
is synchronously updated, the time-dependent change of
$p({\Theta},t)$ is described as follows:
\begin{equation}
  \label{eq:t+1}
    p({\Theta},t+\Delta t)=
\sum_{{\Theta}^{\prime}}
p({\Theta},t+\Delta t|{\Theta}^{\prime},t)
p({\Theta}^{\prime},t) 
\end{equation}
where ${\Theta}^{\prime}$ denotes all possible realizations of
${\Theta}$ and $p({\Theta},t+\Delta
t|{\Theta}^{\prime},t)$ denote the conditional probabilities to go
from state ${\Theta}^{\prime}$ at time $t$ to ${\Theta}$ at
time $t+\Delta t$. \Eqn{eq:t+1} is based on the Markov assumption that
the dynamics at time $t+\Delta t$ may depend only on states at time $t$.
With the assumption of small time steps $\Delta t$ and the definition of
the transition rates
\begin{equation}
  \label{trans0}
w({\Theta}|{\Theta}^{\prime},t) = \lim_{\Delta t \to 0}
\frac{p({\Theta},t+\Delta t|{\Theta}^{\prime},t)}{\Delta t}
\end{equation}
\eqn{eq:t+1} can be transferred into a time-continuous master equation as
follows:
\begin{equation}
\label{master0}
\frac{d}{dt}p({\Theta},t)=\sum_{{\Theta}^{\prime}} \Big[
w({\Theta}|{\Theta}^{\prime})\;
p({\Theta}^{\prime},t) 
- w({\Theta}^{\prime}|{\Theta})\;
p({\Theta},t)\Big]
\end{equation}
In \eqn{master0}, the transition rates depend on the \emph{whole}
distribution ${\Theta}$. However, in the frequency dependent dynamics
introduced in Sect. \ref{2.1}, only the occupation distribution of the
\emph{local} neighborhood of node $i$ needs to be taken into account.
Therefore, it is appropriate to think about some reduced description in
terms of lower order distributions, such as the local occupation
$\ul{\theta}_{i}$, \eqn{occupat}.  In principle, there are two different
ways to solve this task. The first one, the \emph{top-down} approach
starts from the \emph{global} distribution ${\Theta}$ in the whole state
space and then uses different approaches to factorize $p({\Theta},t)$.
However, a Markov analysis \citep{muehlenb-hoens-02} can only be carried
out exactly for small $N$, because of the exponential $N$-dependence of
the state space.  Thus, for larger $N$ suitable approximations, partly
derived from theoretical concepts in computer
science 
need to be taken into account.

In this paper, we follow a second way which is a \emph{bottom-up}
approach based on the \emph{local} description already given in Sect.
\ref{2.1}. I.e. starting from node $i$ and its local neighborhood, we
want to derive the dynamics for some appropriate \emph{macroscopic}
variables describing the nonlinear VM.  Instead of one equation for
$p({\Theta},t)$ in the top-down approach, in the bottom-up approach we
now have a set of $N$ stochastic equations for $p_i(\theta_i,t)$, eqn.
(\ref{master0}), which are locally coupled because of overlapping
neighborhoods, $\ul{\theta}_{i}$. In order to solve the dynamics, we need
to discuss suitable approximations for these local correlations. As we
are interested in the macroscopic dynamics, these approximations will be
done at the macroscopic level.  In order to do so, we first derive a
macroscopic equation from the stochastic \eqn{master}, which is carried
out in the following section.

\subsection{Derivation of the macroscopic dynamics}
\label{3.1}
The key variable of the macroscopic dynamics is the global frequency
$x_{\sigma}(t)$, defined in \eqn{nconst}. In order to compare the
averaged computer simulations with results from analytical approximations
later in \sect{4}, we first derive an equation for the expectation value
$\mean{x_{\sigma}}$. We do this without an explicit determination of the
transition rates and wish to emphasize that the formal approach presented
in Sect. \ref{2.2} remains valid not just for the voter model, but also
for other dynamic processes which depend on neighbor interactions (not
only nearest neighbors) in various network topologies.

For the derivation of the expectation value we start from the stochastic
description given in \sect{sec:micro}, where $p(\Theta,t)$ denoted the
probability to find a particular distribution $\Theta$, \eqn{vector}, at
time $t$ and $\Theta^{\prime}$ denoted all possible realizations of
$\Theta$ \eqn{master0}. On one hand:
\begin{eqnarray}
\label{eq:average}
\mean{x_{\sigma} (t)}&=& \frac{1}{N} \sum_{\Theta^{\prime}}
\left(\sum_{i=1}^{N} \delta_{\sigma\theta_i} \right)
p(\Theta^{\prime},t) \nonumber \\ 
&=& \frac{1}{N} \sum_{\Theta^{\prime}} N_{\sigma}\, p(\Theta^{\prime},t)
=\frac{\mean{N_{\sigma} (t)}}{N} 
\end{eqnarray}
and on the other hand:
\begin{eqnarray}
\label{av2}
\mean{x_{\sigma} (t)} & = & 
\frac{1}{N}\sum_{i=1}^{N} \sum_{\Theta^{\prime}}
\delta_{\sigma\theta_i}\, p(\Theta^{\prime},t) \nonumber \\
&=&\frac{1}{N}\sum_{i=1}^{N}  p_i(\theta_i\=\sigma,t)
\end{eqnarray}
By differentiating \eqn{av2} with respect to time and inserting the
master \eqn{master}, we find the following macroscopic dynamics for
the network:
\begin{eqnarray}
\label{eq:master_macro1}
\frac{d}{dt}\mean{x_{\sigma}(t)}=  
\frac{1}{N}\sum_{i=1}^{N} \sum_{\ul{\theta}_{i}^{\prime}} 
 \Big[ w(\sigma|(1\-\sigma),\ul{\theta}_{i}^{\prime}) 
\times \nonumber  \\ 
\times p(\theta_i\=(1\-\sigma),\ul{\theta}_{i}^{\prime},t) \nonumber \\
- w(1\-\sigma|\sigma,\ul{\theta}_{i}^{\prime}) \,
p(\theta_i\=\sigma,\ul{\theta}_{i}^{\prime},t) \Big]
\end{eqnarray}
For the further treatment of \eqn{eq:master_macro1}, we consider a
specific distribution of states on $m+1$ nodes defined by
$\ul{\sigma}^{0}$. This distribution is assigned to a particular
neighborhood of node $i$ by $\ul{\eta}^{0}_{i}(\ul{\sigma}^{0})$
(\eqn{occupat0}). Since we are interested in how many times a special
realization of a specific distribution $\ul{\sigma}^0$ is present in the
population, we define an indicator function
\begin{equation}
  \label{chi}
\chi(\ul{\eta}_{i}^{0}) \equiv
\chi_{\ul{\sigma}^0}(\ul{\eta}_{i}^{0}(\ul{\sigma}^{0}))=\delta_{\ul{\theta}_i\ul{\sigma}^0}
\end{equation}
that is $1$ if the neighborhood of node $i$ has the distribution
$\ul{\sigma}^0$, and $0$ otherwise. Therefore, we write the frequency of the
$n$-tuplet $\ul{\sigma}^{0}$ in the population as:
\begin{equation}
\label{eq:x_ind}
x_{\ul{\sigma}^{0}}(\Theta) := \frac{1}{N}\sum_{i=1}^{N}
\chi(\ul{\eta}_{i}^{0})
\end{equation}
The expectation value is
\begin{equation}
\label{eq:av_x}
\mean{x_{\ul{\sigma}^{0}}(t)} = \sum_{\Theta^{\prime}}
x_{\ul{\sigma}^{0}}(\Theta^{\prime})\;p(\Theta^{\prime},t)
\end{equation}
Inserting \eqn{eq:x_ind} into \eqn{eq:av_x}, we verify that
\begin{equation}
\label{eq:av_x1}
\begin{aligned}
\mean{x_{\ul{\sigma}^{0}}(t)} =&
\frac{1}{N} \sum_{i=1}^{N} \sum_{\Theta^{\prime}}
\chi(\ul{\eta}_{i}^{0})
\; p(\Theta^{\prime},t) \\
= & \frac{1}{N}\sum_{i=1}^{N} p(\ul{\eta}_{i}^{0},t)
\end{aligned}
\end{equation}
because of the definition of the marginal distribution.  Using the
identity $p(\ul{\eta}^{0}_{i},t)=p(\sigma,\ul{\eta}_{i},t)$, we may
rewrite \eqn{eq:master_macro1} by means of \eqn{eq:av_x1} to derive the
macroscopic dynamics in the final form:
\begin{equation}
\label{eq:master_macrof}
\begin{aligned}
  \frac{d}{dt} \mean{x_{\sigma}(t)} = \sum_{\ul{\sigma}^{\prime}}& \Big[
  w(\sigma|(1\-\sigma),\ul{\sigma}^{\prime})\,
  \mean{x_{(1\-\sigma),\ul{\sigma}^{\prime}}(t)} \\
  & - w(1\-\sigma|\sigma,\ul{\sigma}^{\prime})\,
  \mean{x_{\sigma,\ul{\sigma}^{\prime}}(t)}\Big ]
\end{aligned}
\end{equation}
$\ul{\sigma}^{\prime}$ denotes the $2^{m}$ possible configurations of a
specific occupation distribution $\ul{\sigma}$, \eqn{sigm01}).  In the
following, we use $\mean{x}\equiv\mean{x_{1}}=1-\mean{x_{0}}$. Then, the
dynamic for $\mean{x}$ reads:
\begin{equation}
\label{x-fin}
\begin{aligned}
\frac{d}{dt} \mean{x(t)} =  \sum_{\ul{\sigma}^{\prime}}& \Big[
w(1|0,\ul{\sigma}^{\prime})\,
\mean{x_{0,\ul{\sigma}^{\prime}}(t)} \\
& - w(0|1,\ul{\sigma}^{\prime})\,
\mean{x_{1,\ul{\sigma}^{\prime}}(t)}\Big ]
\end{aligned}
\end{equation}
The solution of \eqn{x-fin} would require the computation of the averaged
global frequencies $\mean{x_{1,\ul{\sigma}}}$ and
$\mean{x_{0,\ul{\sigma}}}$ for all possible occupation patterns
$\ul{\sigma}$, which would be a tremendous effort. Therefore, in the next
section we will introduce two analytical approximations to solve this
problem.  In \sect{4.1} we will further show by means of computer
simulations that these approximations are able to describe the averaged
dynamics of the nonlinear VM.

\subsection{Mean-Field Limit and Pair Approximation}
\label{3.2}

As a first approximation of eqn. (\ref{x-fin}), we investigate the
mean-field limit. Here the state of each node does not depend on the
occupation distribution of its neighbors, but on $m$ randomly chosen
nodes. In this case the occupation distribution factorizes:
\begin{equation}
  \label{factorize}
\mean{x_{\ul{\sigma}^{0}}}= \mean{x_{\sigma}}\, \prod_{j=1}^{m}
\mean{x_{\sigma_j}} 
\end{equation}
For the macroscopic dynamics, \eqn{x-fin}, we find:
\begin{equation}
\label{eq:master_macro_mean}
\begin{aligned}
\frac{d}{dt}\mean{x(t)}= \sum_{\ul{\sigma}^{\prime}}\Big[&
w(1|0,\ul{\sigma}^{\prime})\, (1-\mean{x}) \prod_{j=1}^{m}
\mean{x_{\sigma_j}} \\ & - 
w(0|1,\ul{\sigma}^{\prime}) \,\mean{x} \prod_{j=1}^{m}
\mean{x_{\sigma_j}}\Big]
\end{aligned}
\end{equation}
For the calculation of the $\mean{x}_{\sigma_{j}}$ we have to look at
each possible occupation pattern $\ul{\sigma}$ for a neighborhood $m$.
This will be done in detail in Sect. \ref{sec:a1a2}. Before, we discuss
another analytical approximation which solves the macroscopic \eqn{x-fin}
with respect to \emph{correlations}.  This is the so-called \emph{pair
  approximation}, where one is not interested in the occupation
distribution of a whole neighborhood $\ul{\sigma}^{0}$, \eqn{occupat0})
but only in \emph{pairs} of nearest neigbor nodes,
$\sigma,\sigma^{\prime}$ with $\sigma^{\prime}\in\{0,1\}$. That means the
local neighborhood of nearest neighbors is decomposed into pairs, i.e.
blocks of size 2 that are called \emph{doublets}.
 
Similar to \eqn{eq:x_ind}, the global frequency of doublets is defined
as:
\begin{equation}
\label{d_den}
x_{\sigma,\sigma^{\prime}}=\frac{1}{N} \sum_{i=1}^{N}
\sum_{j=1}^{m} \frac{1}{m} \;
\chi(\theta_i=\sigma,\theta_{i_j}=\sigma^{\prime})
\end{equation}
The expected value of the doublet frequency is then given by
$\mean{x_{\sigma,\sigma^{\prime}}}$ in the same way as in \eqn{eq:av_x}.
We now define the correlation term as:
\begin{equation}
\label{eq:c-prob}
c_{\sigma|\sigma^{\prime}}:=\frac{\mean{x_{\sigma,\sigma^{\prime}}}}{\mean{x_{\sigma^{\prime}}}}
\end{equation}
neglecting higher order correlations.  Thus $c_{\sigma|\sigma^{\prime}}$
can be seen as an approximation of the conditional probability that a
randomly chosen nearest neighbor of a node in state $\sigma^{\prime}$ is
in state $\sigma$.  Using the above definitions, we have the following
relations:
\begin{equation}
\label{eq:cond_den}
\mean{x_{\sigma^{\prime}}}c_{\sigma|\sigma^{\prime}}=\mean{x_{\sigma}}
 c_{\sigma^{\prime}|\sigma} \;;
\sum_{\sigma^{\prime} \in \{0,1\}} c_{\sigma^{\prime}|\sigma}=1 
\end{equation}
For the case of two species $\sigma\in\{0,1\}$, $c_{1|1}$ and $c_{0|0}$ are
the \emph{inter-species} correlations, while $c_{1|0}$ and $c_{0|1}$ denote
the \emph{intra-species} correlations. Using $\mean{x}\equiv\mean{x_{1}}$,
these correlations can be expressed in terms of only $c_{1|1}$ and $\mean{x}$
as follows:
\begin{equation}
\label{eq:cond_prob}
\begin{aligned}
c_{0|1}=& 1-c_{1|1}\\
c_{1|0}=& \frac{\mean{x}\, (1-c_{1|1})}{1-\mean{x}}\\
c_{0|0}=& \frac{1-2\mean{x} +\mean{x}c_{1|1}}{1-\mean{x}}
\end{aligned}
\end{equation}
Now, the objective is to express the global frequency of a specific
occupation pattern $\mean{x_{\ul{\sigma}^{0}}}$, \eqn{eq:av_x}, in terms
of the correlation terms $c_{\sigma|\sigma^{\prime}}$. In pair
approximation, it is assumed that the states $\theta_{i_j}$ are
correlated only through the state $\theta_i$ and uncorrelated otherwise.
Then the global frequency terms in \eqn{eq:master_macrof} can be
approximated as follows:
\begin{equation}
\label{eq:pairapprox}
\mean{x_{\ul{\sigma}^{0}}} =\mean{x_{\sigma}} \prod_{j=1}^{m}
  c_{{\sigma_j}|\sigma} 
\end{equation}
For the macroscopic dynamics, \eqn{x-fin}, we find in pair approximation:
\begin{equation}
\label{macro-pair}
\begin{aligned}
\frac{d}{dt}\mean{x(t)}= \sum_{\ul{\sigma}^{\prime}}\Big[&
w(1|0,\ul{\sigma}^{\prime})\, (1-\mean{x}) \prod_{j=1}^{m}
c_{\sigma_{j}|\sigma} \\ & - 
w(0|1,\ul{\sigma}^{\prime}) \,\mean{x} \prod_{j=1}^{m}
c_{\sigma_{j}|(1-\sigma)} \Big]
\end{aligned}
\end{equation}
Note that the $c_{\sigma_{j}|\sigma}$ can be expressed in terms of
$c_{1|1}$ by means of \eqn{eq:cond_prob}. Thus, \eqn{macro-pair} now
depends on only two variables, $\mean{x}$ and $c_{1|1}$.
In order to derive a \emph{closed} form description, we need an additional
equation for $\dot{c}_{1|1}$. That can be obtained from \eqn{eq:c-prob}:
\begin{equation}
\label{eq:local1}
\frac{dc_{1|1}}{dt} =  - \frac{c_{1|1}}{\mean{ x}}
\frac{d}{dt}\mean{ x}+\frac{1}{\mean{ x}}
\frac{d}{dt}\mean{ x_{1,1} }
\end{equation}
\Eqn{eq:local1} also requires the time derivative of the global doublet
frequency $\mean{x_{1,1}}$. Even in their lengthy form, the three
equations for $\mean{ x}$, $c_{1|1}$, $\mean{ x_{1,1} }$
can easily be solved numerically. This gives the approach some
computational advantage compared to averaging over a number of
microscopic computer simulations for all possible parameter sets.  

Although the approach derived so far is quite general in that it can be
applied to different network topologies and neighborhood sizes, specific
expressions for these three equations of course depend on
these. Therefore, in the Appendix, these three equations are explicitly
derived for a 2d regular lattice with neighborhood $m=4$ using the
specific transition rates introduced in the next section. In
Sect. \ref{4.1}, we further show that the pair approximation yields some
characteristic quantities such as $\mean{x(t)}$ for the 2d regular
lattice in very good agreement with the results of computer simulations.

\section{Invasion versus Coexistence}
\label{2.3}

\subsection{Nonlinear Response Functions}
\label{sec:a1a2}

So far, we have developed a stochastic framework for (but not restricted
to) nonlinear voter models in a general way, without specifying two of
the most important features, namely (i) the network topology which
defines the neighborhood, and (ii) the nonlinearity $\kappa(f)$ which
defines the response to the local frequencies of the two different
states. For (i), let us choose a regular network with $m=4$, i.e. each
voter has 4 different neighbors.  We note explicitly that our modeling
framework and the general results derived hold for \emph{all homogeneous
  networks}, but for the visualization of the results it will be most
convenient to choose a regular square lattice, where the neighbors appear
next to a node. This allows us to observe the formation of macroscopic
ordered states in a more convenient way, without restricting the general
case.  Eventually, to illustrate the dynamics let us now assume a
population biology context, where each node is occupied by an individual
of either species 0 or 1. The spreading of one particular state is then
interpreted as the invasion of that respective species and the local
disappearence of the other one, while the emergence of a complete ordered
state is seen as the complete invasion or domination of one species
together with the extinction of the other one.

Keeping in mind that we also consider the state of node $i$ itself, we
can write the possible transition rates, eqn. (\ref{eq:nonlinvm}), for
the neighborhood of $n=m+1=5$ and $\theta_{i}=\sigma$ in the following
explicit way (cf also \citep{molofsky99}):
\begin{equation}
\begin{array}{ccc}
        \quad f_{i}^{\sigma} \quad &\quad f_{i}^{(1\-\sigma)} \quad&
        w(1\-\theta_{i}|\theta_{i}\=\sigma,f_{i}^{\sigma}) \\ \hline
         1& 0 & \alpha_{0}\\
         4/5&1/5&\alpha_1\\
         3/5&2/5&\alpha_2\\
         2/5&3/5&\alpha_{3}=1\-\alpha_2\\
         1/5&4/5&\alpha_{4}=1\-\alpha_1\\
         0&5/5&\alpha_{5}=1\-\alpha_0
        \end{array}
        \label{trans2}
\end{equation}
\Eqn{trans2} means that a particular node $i$ currently in state
$\theta_{i}=\sigma$, or occupied by an individual of species $\sigma$
where $\sigma$ is either 0 or 1, will be occupied by an individual of
species $(1\-\sigma)$ with a rate
$w(1\-\theta_{i}|\theta_{i}\=\sigma,f_{i}^{\sigma})$ that changes with
the local frequency $f_{i}^{\sigma}$ in a \emph{nonlinear} manner.  The
different values of $\alpha_{n}$ denote the products
$\kappa(f)f_{i}^{1-\sigma}$ for the specific values of $f$ given. I.e.,
the $\alpha_{n}$ define the piecewise linear functions shown in
Fig. \ref{fig:nonlin}. 

The general case of six independent transition rates $\alpha_{n}$
(n\=0,...,5) in \eqn{trans2} can be reduced to three transition rates
$\alpha_{0}$, $\alpha_{1}$, $\alpha_{2}$ by assuming a symmetry of the
invasion dynamics of the two species, i.e. $\alpha_{2}+\alpha_{3}=1$,
$\alpha_{1}+\alpha_{4}=1$ and $\alpha_{0}+\alpha_{5}=1$. Further,
assuming a pure frequency dependent process, we have to consequently
choose $\alpha_{0}=0$, because in a complete homogeneous neighborhood,
there is no incentive to change to another state (there are no other
species around to invade).

\begin{figure}[htbp]
  \begin{center}
{\psset{unit=5}
    \begin{pspicture}(1,1)
\pspolygon[fillcolor=yellow,fillstyle=solid,linewidth=1pt,linestyle=solid]
(0,0.5)(0.5,0.5)(1,1)(0,1)(0,0.5)
\pspolygon[fillcolor=green,fillstyle=solid,linewidth=1pt,linestyle=solid]
(0,0)(0.5,0.5)(1,0.5)(1,0)(0,0)
\pspolygon[fillcolor=white,fillstyle=solid,linewidth=1pt,linestyle=solid]
(0.,0.)(0.,0.5)(0.5,0.5)(0,0.)
\pspolygon[fillcolor=lightgray,fillstyle=solid,linewidth=1pt,linestyle=solid]
(0.5,0.5)(1,0.5)(1,1)(0.5,0.5)
\pspolygon[linewidth=1pt](0,0)(0,1)(1,1)(1,0)(0,0)
\psaxes[Dx=0.2,Dy=0.2,axesstyle=frame](0,0)(1,1)
\rput(0.2,0.4){{\Large $\bullet$}}
\rput(0.13,0.30){\Large \textbf{(pf)}}
\rput(0.85,0.65){\Large \textbf{(nf)}}
\rput(0.3,0.75){\Large \textbf{(pa)}}
\rput(0.7,0.2){\Large \textbf{(na)}}
\rput(0.5,-0.2){\Large $\alpha_{1}$}
\rput{90}(-0.2,0.5){\Large $\alpha_{2}$}
    \end{pspicture}
}
\vspace*{8mm}
  \end{center}
\caption[]{
  Four different parameter regions for frequency dependent invasion,
  according to \eqn{eps-alpha}. The linear voter point is indicated by
  $\bullet$ \label{region}}
\end{figure}
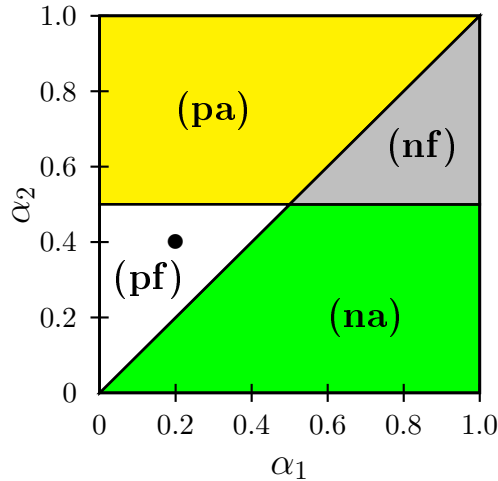

We recall that if the transition rates $\alpha_{1}$, $\alpha_{2}$ are
directly proportional to $f^{(1-\sigma)}$, i.e. $\alpha_{1}=0.2$ and
$\alpha_{2}=0.4$, this recovers the linear VM, eqn. (\ref{eq:linvm}).
(Note that without the resistence of node $i$ discussed in Sect.
\ref{sec:vm} the linear voter point would read as $\alpha_{1}=0.25$ and
$\alpha_{2}=0.5$ instead.)  Dependent on the relation of the two
essential parameters $\alpha_{1}$, $\alpha_{2}$, we also find different
versions of \emph{nonlinear} VM, which have their specific meaning in a
population biology context:
\begin{eqnarray}
  \label{eps-alpha}
\mathrm{(pf)} \quad &
0 \leq \alpha_{1} \leq \alpha_{2}\leq (1-\alpha_{2}) \leq
(1-\alpha_{1}) \leq1  \nonumber \\
\mathrm{(nf)}  \quad &
1 \geq \alpha_{1} \geq \alpha_{2}\geq (1-\alpha_{2}) \geq
(1-\alpha_{1}) \geq 0 \nonumber \\  
\mathrm{(pa)}  \quad &
0 \leq \alpha_{1} \leq \alpha_{2}, \; \alpha_{2} \geq (1-\alpha_{2}), 
\nonumber\\
 & (1-\alpha_{2}) \leq
(1-\alpha_{1}) \leq1  \nonumber \\
\mathrm{(na)}  \quad &
1 \geq \alpha_{1} \geq \alpha_{2}, \; \alpha_{2} \leq (1-\alpha_{2}), 
\nonumber \\
&  (1-\alpha_{2})\geq
(1-\alpha_{1}) \geq 0  
\end{eqnarray}
Note, that the parameters $\alpha_1,\dots,\alpha_4$ can be ordered in
$24$ different ways. These reduce to $8$ inequalities under the
conditions $\alpha_3=1-\alpha_2$ and $\alpha_4=1-\alpha_1$.  In
\eqn{eps-alpha}, (pf) means (pure) \emph{positive freqency dependent
  invasion}, where the transition rate \emph{increases} with an
increasing number of individuals of the \emph{opposite} species
$(1\-\sigma)$ in the neighborhood, and (nf) means (pure) \emph{negative
  freqency dependent invasion} because the transition rate
\emph{decreases}.  The two other cases describe positive (pa) and
negative (na) \emph{allee effects} \citep{molofsky99}.  These regions
are described by $3$ inequalities each, all of which show the same
relative change in parameter values, if going from $\alpha_1$ to
$\alpha_4$. Similar to the drawings in Fig. \ref{fig:nonlin} this can be
roughly visualized as an up-down-up change in region (pa) and a
down-up-down change in region (na). The different parameter regions are
shown in Fig.  \ref{region}. On a first glimpse, one would expect that
the dynamics as well as the evolution of global variables may be
different in these regions. Thus, one of the aims of this paper is to
investigate whether or to what extent this would be the case.

\subsection{Mean-Field Analysis}
\label{3}

In order to find out about the influence of the nonlinear response
function $\kappa(f)$, which is specified here in terms of $\alpha_{1}$,
$\alpha_{2}$, let us start with the mean-field approach that lead to eqn.
(\ref{eq:master_macro_mean}). As we outlined in Sect.  \ref{3.2}, the
calculation of the $\mean{x_{\sigma_{j}}}$ in eqn.
(\ref{eq:master_macro_mean}) requires to look at each possible occupation
pattern $\ul{\sigma}$ for a neighborhood $m$, for instance,
$\ul{\sigma}=\{0010\}$. The mean-field approach assumes that the
occurence of each 1 or 0 in the pattern can be described by the global
frequencies $x$ and $(1\-x)$, respectively (for simplicity, the
abbreviation $x\equiv\mean{x}$ will be used in the following). For the
example of string $\ul{\sigma}=\{0010\}$ we find $\prod
\mean{x_{\sigma_{j}}}= x(1\-x)^{3}$.  The same result yields for
$\ul{\sigma}=\{0100\}$ and for any other string that contains the same
number of 1 and 0, i.e. there are exactly ${4 \choose 1}$ different
possibilities. For strings with two nodes of each species, ${4 \choose
  2}$ times the contribution $\prod \mean{x_{\sigma_{j}}}=
x^{2}(1\-x)^{2}$ results, etc. Inserting \eqn{trans2} for the transition
rates, we find with $\alpha_{0}=0$ the equation for the mean-field
dynamics:
\begin{equation}
\label{eq:mean_field}
\begin{aligned}
  \frac{dx}{dt} = \; (1-x) &
  \Big[ 
4 \alpha_{1}  x(1-x)^{3} +  6 \alpha_{2} x^{2}(1-x)^{2} \\
  & + 4 (1-\alpha_{2}) x^{3} (1-x) +  (1-\alpha_{1}) x^{4} \Big ] \\
  \; - x & \Big[ (1-\alpha_{1}) (1-x)^{4} + 4 (1-\alpha_{1}) x(1-x)^{3}  \\ 
& + 6 \alpha_{2}  x^{2}(1-x)^{2} + 4 \alpha_{1} x^{3} (1-x) 
\Big]
\end{aligned}
\end{equation}
The fixed points of the mean-field dynamics can be calculated from
\eqn{eq:mean_field} using $\dot{x}=0$.  We find:
\begin{eqnarray}
  \label{zero}
   x^{(1)} &=& 0\;;\quad x^{(2)}= 1\;;\quad x^{(3)}=0.5 \nonumber \\
x^{(4,5)} & =& 0.5 \pm \sqrt{\beta_{1}/4\beta_{2}} \\
  \beta_{1}& \equiv& \alpha_{2} + 1.5\alpha_{1} - 0.7 ;\; 
\beta_{2}\equiv \alpha_{2}-0.5 \alpha_{1}-0.3 \nonumber
\end{eqnarray}
The first three stationary solutions denote either a complete invasion
of one species or an equal share of both of them. ``Nontrivial''
solutions, i.e.  a \emph{coexistence} of both species with different
shares of the total population, can only result from $x^{(4,5)}$,
provided that the solutions are (i) real and (ii) in the interval
$\{0,1\}$. The first requirement means that the two functions
$\beta_{1}$, $\beta_{2}$ are either both positive or both negative. The
second requirement additionally results in $\alpha_{1}\leq 0.2$ if
$\alpha_{2}\geq 0.4$ and $\alpha_{1}\geq 0.2$ if $\alpha_{2}\leq 0.4$.
This leads to the phase diagram of the mean-field case shown in
\pic{phase}.
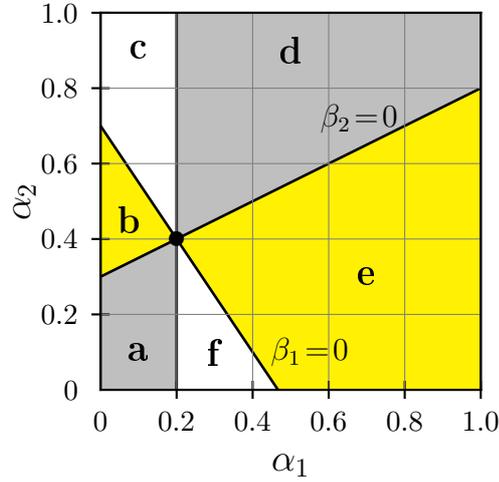
\begin{figure}[htbp]
  \begin{center}
{\psset{unit=5}
    \begin{pspicture}(1,1)
\pspolygon[fillcolor=lightgray,fillstyle=solid,linestyle=none]
(0,0)(0.2,0)(0.2,0.4)(0,0.3)(0,0)
\pspolygon[fillcolor=lightgray,fillstyle=solid,linestyle=none]
(0.2,1)(0.2,0.4)(1,0.8)(1,1)(0.2,1)
\pspolygon[fillcolor=yellow,fillstyle=solid,linestyle=none]
(0.2,0.4)(0.4667,0)(1,0)(1,0.8)(0.2,0.4)
\pspolygon[fillcolor=yellow,fillstyle=solid,linestyle=none]
(0.2,0.4)(0,0.3)(0,0.7)(0.2,0.4)
\psline[linewidth=1pt,linestyle=solid](0,0.7)(0.4667,0.0)
\psline[linewidth=1pt,linestyle=solid](0,0.3)(1,0.8)
\psline[linewidth=1pt,linestyle=solid](0.2,0)(0.2,1)
\psaxes[Dx=0.2,Dy=0.2,axesstyle=frame](0,0)(1,1)
\psgrid[gridlabels=0]
\rput(0.2,0.4){{\Large $\bullet$}}
\rput(0.1,0.1){\Large \textbf{a}}
\rput(0.075,0.45){\Large \textbf{b}}
\rput(0.1,0.9){\Large \textbf{c}}
\rput(0.5,0.9){\Large \textbf{d}}
\rput(0.7,0.3){\Large \textbf{e}}
\rput(0.3,0.1){\Large \textbf{f}}
\rput(0.5,-0.2){\Large $\alpha_{1}$}
\rput{90}(-0.2,0.5){\Large $\alpha_{2}$}
\rput(0.68,0.72){\large $\beta_{2}\!=\! 0$}
\rput(0.55,0.1){\large $\beta_{1}\!=\! 0$}
    \end{pspicture}
  }
\vspace*{8mm}
\end{center}
\caption[]{Phase diagram of the invasion dynamics in the mean-field
  case.  For the different areas see text.  The functions $\beta_{1}=0$
  and $\beta_{2}=0$ are given by \eqn{zero}.  The areas in lighter gray
  indicate imaginary solutions of $x^{(4,5)}$, \eqn{zero} while the areas
  in darker gray (color: yellow) indicate solutions of $x^{(4,5)}$
  outside the $\{0,1\}$  interval. The linear voter point is indicated by
  $\bullet$. Note that the physically relevant solutions $x^{1,2,3}$ are
  the same in the lighther gray and darker gray (color: yellow) areas,
  however their stability is different in (a,b) and (d,e).  
  \label{phase}}
\end{figure}
\begin{figure}[htbp]
  \begin{center}
  \includegraphics[width=6.5cm]{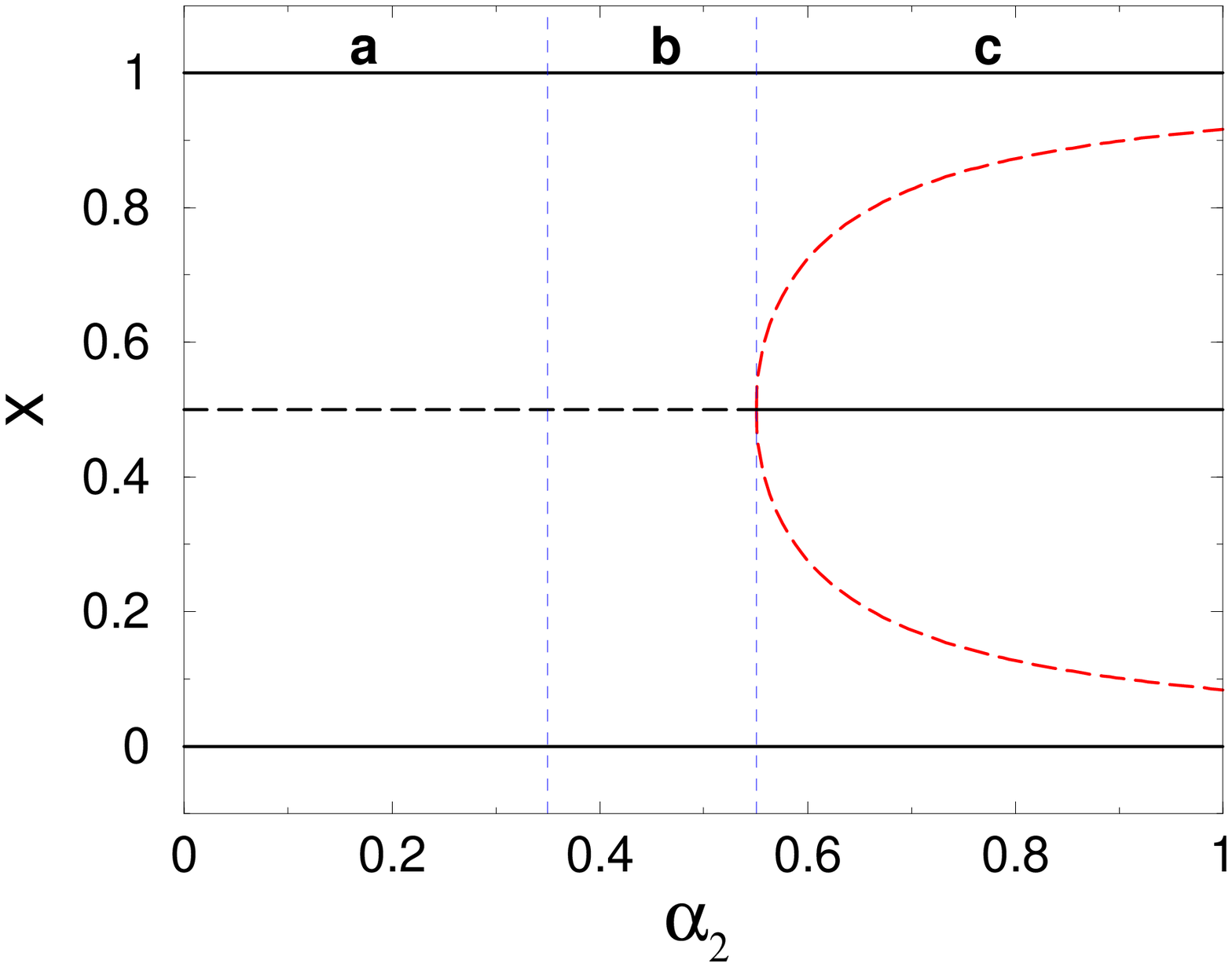}\hfill
 \includegraphics[width=6.5cm]{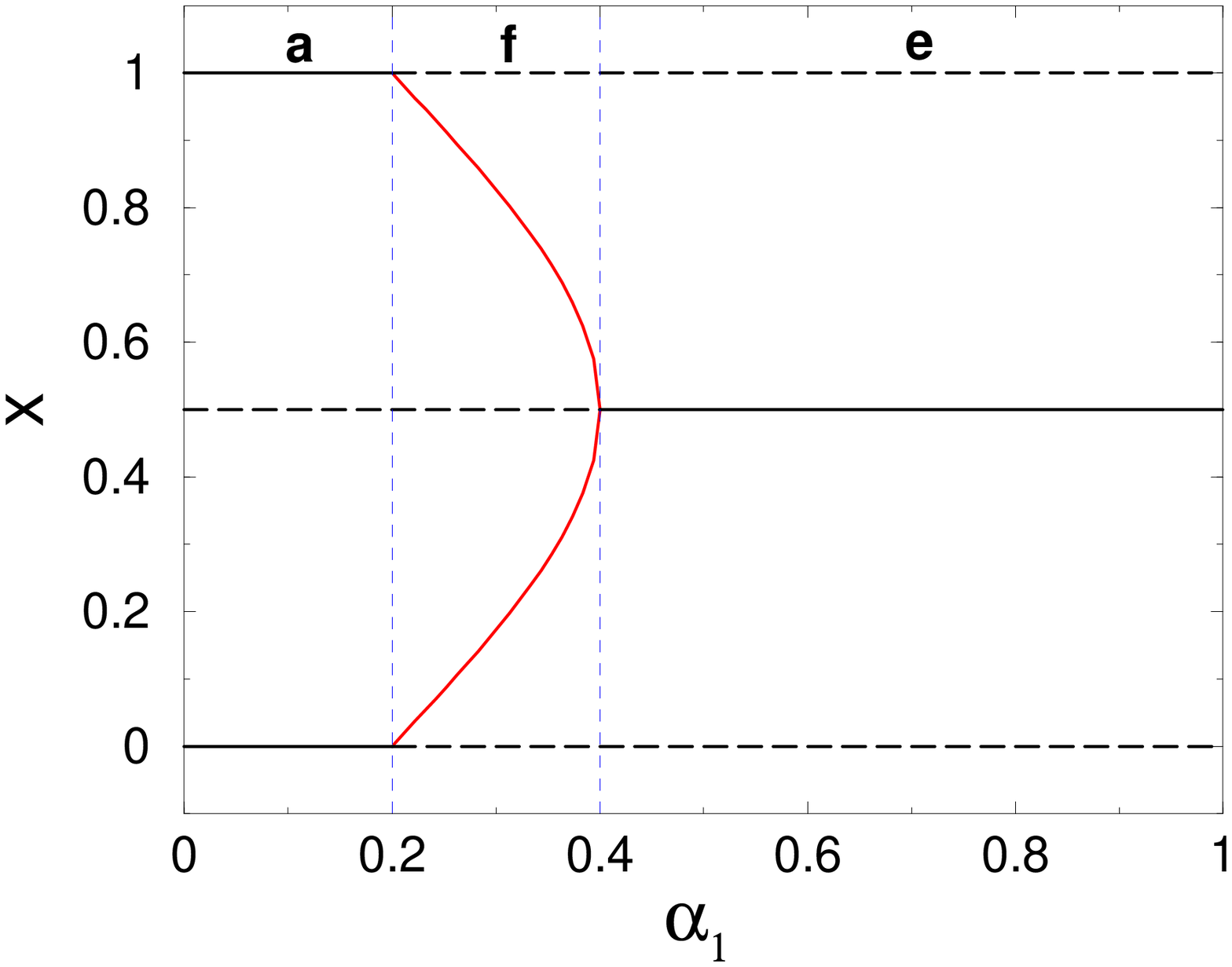}
  \end{center}
    \caption[]{
      Bifurcation diagram of the stationary solutions dependent on
      $\alpha_{1}$ and $\alpha_{2}$. (top) $\alpha_{1}=0.1$, (bottom)
      $\alpha_{2}=0.1$. The
      solid lines refer to stable solutions, the dashed lines to unstable
      ones. The notations a-f refer to the respective areas in the phase
      diagram, \pic{phase}. 
      \label{a1a2}}
  \end{figure}

In order to verify the stability of the solutions, we have further done a
perturbation analysis (see also Sect. \ref{sec:pertub}).  
The results can be summarized as follows:
\begin{itemize}
\item In the regions $a$ and $b$ of the mean-field phase diagram,
  \pic{phase}, $x^{\mathrm{stat}}=0$ and $x^{\mathrm{stat}}=1$ are the
  only stable fixed points of the dynamics, while $x=0.5$ is an unstable
  fixed point (cf. also \pic{a1a2}top). Species 1 with $x(t\=0)<0.5$ will
  most likely become extinct, while it will remain as the only survivor
  for $x(t\=0)>0.5$. Thus, the region $(a,b)$ can be characterized as
  the region of \emph{invasion}.
\item In region $c$, the mean-field limit predicts the three stable fixed
  points $0$, $1$ and $0.5$. The attractor basin for $0.5$ is the largest
  as \pic{a1a2}(top) indicates. The separatices are given by the unstable
  solutions $x^{(4,5)}$, \eqn{zero}.  In this parameter region, the
  mean-field limit predicts either \emph{coexistence} of both species
  with equal shares, or \emph{invasion} of one species, dependent on the
  initial condition $x(t\=0)$.
\item In the regions $d$ and $e$, only one stable fixed point
  $x^{\mathrm{stat}}=0.5$ can be found, while the solutions $0$ and $1$
  are unstable (cf. also \pic{a1a2}bottom). Thus, the mean-field approach
  predicts the \emph{coexistence} of both species with equal share.
\item Finally, in region $f$ the solutions 0, 1 and $0.5$ are unstable
  fixed points , but the two remaining solutions $x^{(4,5)}$, \eqn{zero}
  are stable fixed points (cf. \pic{a1a2}bottom). Thus, this region is
  the most interesting one, since it seems to enable ``nontrivial''
  solutions, i.e. an \emph{asymmetric coexistence} of both species with
  different shares. We note again, that this is a prediction of the
  mean-field analysis. At the intersection of regions $f$ and $a$, these
  two solutions approach $0$ and $1$, while at the intersection of
  regions $f$ and $e$ they both converge to $0.5$.
\end{itemize}
We will compare these mean-field predictions both with computer
simulations and analytical results from the pair approximations later in
this paper.  Before, in Sects. \ref{sec:limit}, \ref{sec:pertub} we would
like to point to some interesting $(\alpha_{1},\alpha_{2})$ combinations
in this phase diagram where the mean-field analysis does not give a clear
picture of the dynamics.

\subsection{Deterministic limit}
\label{sec:limit}

The first set of interesting points are $(\alpha_{1},\alpha_{2})$
combinations of values 0 and 1, such as $(\alpha_{1},\alpha_{2})=(0,0)$
etc. These cases are special in the sense that they describe the
\emph{deterministic} limit of the nonlinear voter dynamics. Whereas for
$0<\alpha_{n}<1$ always a finite probability exist to change to the
opposite state, for $(0,0)$ the state of node $i$ \emph{never} changes as
long as at least half of the nearest neighbor nodes are occupied by the
same species. On the other hand, it will \emph{always} change if more
then half of the neighboring nodes are occupied by the \emph{other}
species. This refers to a \emph{deterministic} positive frequency
invasion process.  Similarly, a \emph{deterministic} negative frequency
invasion process is described by $(1,1)$.

The deterministic dynamics, as we know from various other examples, may
lead to a completely different outcome as the stochastic counterpart. In
order to verify that we have conducted computer simulations using a
\emph{cellular automaton} (CA), i.e., a two-dimensional regular lattice
with periodic boundary conditions and \emph{synchronous update} of the
nodes. The latter one can be argued, but we verified that there are no
changes in the results of the computer simulations if the sequential
update is used. The time scale for the synchronous update is defined by
the number of simulation steps. If not stated otherwise, the initial
configuration is taken to be a random distribution (within reasonable
limits) of both species, i.e. initially each node is randomly assigned
one of the possible states, $\{0,1\}$. Thus, the initial global frequency
is $x(t\=0)=0.5$.
\pic{determ} shows snapshots of computer simulations of the
deterministic dynamics taken in the (quasi-)stationary dynamic regime.
\begin{figure}[htbp]
\centerline{
\subfigure[$(0,1)$]{
\fbox{\includegraphics[width=3.5cm]{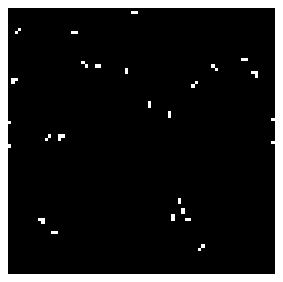}}
\label{det01}}\hspace{2cm}
\subfigure[$(1,1)$]{
\fbox{\includegraphics[width=3.5cm]{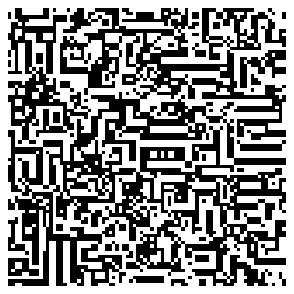}}
\label{det11}}
}
\bigskip
\centerline{
\subfigure[$(0,0)$]{
\fbox{\includegraphics[width=3.5cm]{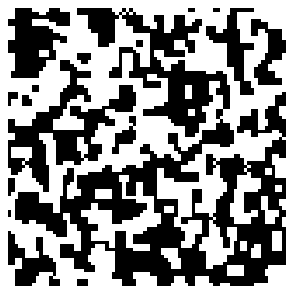}}
\label{det00}}\hspace{2cm}
\subfigure[$(1,0)$]{
\fbox{\includegraphics[width=3.5cm]{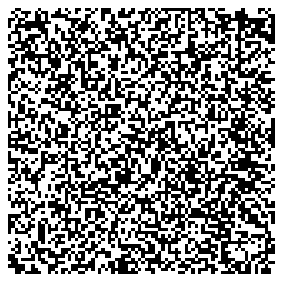}}
\label{det10}}
}
    \caption{Spatial snapshots of a deterministic frequency dependent 
      invasion process for different combinations of
      $(\alpha_{1},\alpha_{2})$. The pictures are taken after $t=10^{2}$
      time steps, except for (a), where $t=10^{4}$. 
      Lattice size: $80\times80$. In all pictures black dots refer to
      species 1.}
    \label{determ}
\end{figure}

If we compare the snapshots of the \emph{deterministic} voter dynamics
with the mean-field prediction, the following observations can be made:
\begin{enumerate}
\item A spatial coexistence of both species is observed for the
  $(\alpha_{1},\alpha_{2})$ values $(0,0)$, Fig. \ref{det00}, $(1,0)$,
  Fig. \ref{det10}, $(1,1)$, Fig. \ref{det11}, where the global frequency
  in the stationary state is $x^{\mathrm{stat}}=0.5$. This contradicts
  with the mean-field prediction for $(0,0)$, which is part of region (a)
  and thus should display complete invasion.

\item A complete invasion of one species is observed for \linebreak
  $(\alpha_{1},\alpha_{2})=(0,1)$, Fig. \ref{det01}. This would agree
  with the mean-field prediction of \emph{either} coexistence \emph{or}
  invasion. A closer look at the bifurcation diagram,
  Fig. \ref{a1a2}(top), however tells us that for the given intial
  condition $x(t=0)=0.5$ the stationary outcome should be
  \emph{coexistence}, whereas the deterministic limit shows \emph{always
    invasion} as was verified by numerous computer simulations with
  varying initial conditions.

\item For the deterministic frequency dependent processes $(0,0)$,
  $(1,0)$ the spatial pattern becomes stationary after a short time. For
  the negative frequency dependence $(1,1)$ the pattern flips between two
  different configurations at every time step. So, despite a constant
  global frequency $x^{\mathrm{stat}}=0.5$ in the latter case, local
  reconfigurations prevent the pattern from reaching a completely
  stationary state, but it may be regarded as (quasi-)stationary, i.e.
  the macroscopic observables do not change but microscopic changes still
  occur.

\item In both -- positive and negative -- deterministic frequency
  dependence cases, individuals of the same species tend to aggregate in
  space, albeit in different local patterns. For the positive frequency
  dependence, we see the occurence of small clusters, Fig. \ref{det00},
  or even complete invasion, Fig. \ref{det01}, based on the local
  feedback within the same species. For the negative frequency dependence
  however we observe the formation of a meander-like structure that is
  also known from physico-chemical structure formation
  \citep{agentbook-03}. It results from the antagonistic effort of each
  species to avoid individuals of the same kind, when being surrounded by
  a majority of individuals of the opposite species.
\end{enumerate}

In conclusion, the mean-field analysis given in this section may provide
a first indication of how the nonlinearities may influence the voter
dynamics. This, however, cannot fully extended to the limiting cases
given by the deterministic dynamics.

\subsection{Perturbation Analysis of the Linear Voter Point}
\label{sec:pertub}

The other interesting $(\alpha_{1},\alpha_{2})$ combination is the linear
voter point $(0.2,0.4)$ where \emph{all} different regions of the
mean-field phase diagram, Fig. \ref{phase}, intersect. Inserting
$(0.2,0.4)$ into eqn.  (\ref{eq:mean_field}) yields $dx/dt=0$ regardless
of the value of $x$, i.e.  $x(t)=x(t=0)$ for \emph{all} initial
conditions. This important feature of the linear VM was already discussed
in Sect.  \ref{sec:vm}. We recall that, while on the one hand the
microscopic realizations always reach consensus (complete invasion of one
species) in the long term, on the other hand an averaged outcome over
many realizations shows that the share of the winning species is
distributed as $x(t=0)$. To put it the other way round, the mean-field
limit discussed failes here because it does not give us any indication of
the fact that there is a completely ordered state in the linear VM.  The
averaged outcome, for example $x=0.5$, can result both from complete
\emph{invasion} of species 1 (50\
\emph{coexistence} of the two species. Both of these outcomes exist in
the immediate neighborhood of the linear voter point as Fig. \ref{phase}
shows.  In order to get more insight into this, we will later use the
pair approximation derived in Sect.  \ref{3.2}.
Here, we first follow a perturbation approach, i.e. we add a small
perturbation to the solution describing the macroscopic ordered state of
complete invasion (consensus). In terms of the nonlinear response
function $\kappa(f)$, expressed by the $\alpha_{n}$ in eqn.
(\ref{trans2}), this means a nonzero value of $\alpha_{0}=\epsilon$,
i.e. a small parameter indicating the perturbation. With this, we arrive
at a modified mean-field equation:
\begin{equation}
\label{eq:mean_field-p}
\frac{dx_{p}}{dt} =
\epsilon \Big[ (1-x)^5 - {x}^5 \Big]+ \frac{dx}{dt} 
\end{equation}
where ${dx}/{dt}$ is given by the nonperturbated mean-field eqn.
(\ref{eq:mean_field}) and the index $p$ shall indicate the presence of
the perturbation $\epsilon$. Consequently, this changes both the value of
the fixed points, previously given by eqn. (\ref{zero}) and their
stability. Instead of a complete analysis in the
$(\epsilon,\alpha_{1},\alpha_{2})$ parameter state, we restrict the
investigations to the vicinity of the linear voter point
$(\alpha_{1},\alpha_{2})=(0.2,0.4)$ where $dx/dt=0$.
Eqn. (\ref{eq:mean_field-p}) then returns only one real stationary
solution, $x_{p}^{stat}=0.5$, which is independent of $\epsilon$ and
stable.  Consequently, any finite perturbation will destroy the
characteristic feature of reaching an ordered state in the linear VM,
i.e. complete invasion, and leaves only \emph{coexistence} of both
species as a possible outcome. This is little surpising because adding an
$\alpha_{0}>0$ to the dynamics transforms the former attractor $x\to0,1$
into a repellor, i.e. it prevents reaching the ordered state.  More
interesting the question is, how the perturbated linear voter dynamics
looks in detail. This is investigated in the next section by means of
computer simulation, and in Sect. \ref{4.1} by means of the pair approximation
approach.

\subsection{Computer simulations of the perturbated CA}
\label{2.4}

For further insight into the dynamics of the nonlinear VM, we perform
some computer simulations using the CA approach already described in
Sect. \ref{sec:limit}. Is important to notice that we have chosen
different sets of the parameters $\alpha_{1}$, $\alpha_{2}$ from the
region of \emph{positive frequency dependence}, as defined in
Fig. \ref{region}. I.e., the transition towards the opposite state
strictly increases with the number of neighbors in that state
(\emph{majority voting}). So one would naively expect a similar
macroscopic dynamics in that region as done in \citep{molofsky99} This
however is not the case as the following simulations indicate. A
thorough analysis is presented in Sect. \ref{4.3}

\begin{figure}[htbp]
\centerline{
\fbox{\includegraphics[width=3.5cm]{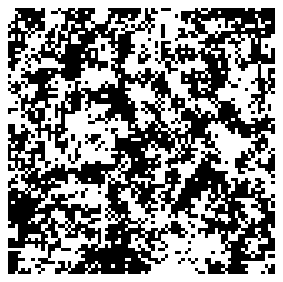}}(a)\hspace{2cm}(b) 
\fbox{\includegraphics[width=3.5cm]{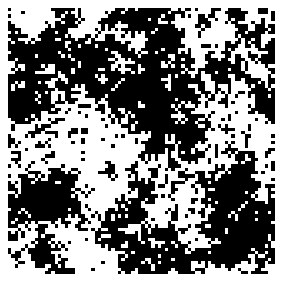}}}
\bigskip
\centerline{
\fbox{\includegraphics[width=3.5cm]{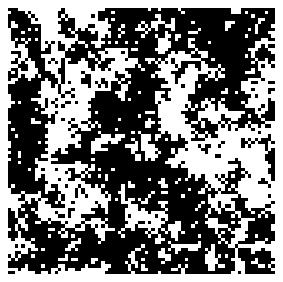}}(c)\hspace{2cm}(d) 
\fbox{\includegraphics[width=3.5cm]{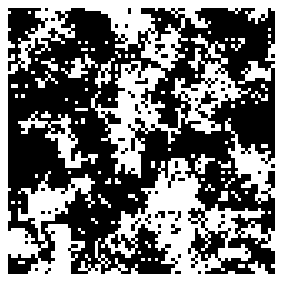}}}
\caption{Spatial snapshots of a positive frequency dependent invasion
  process with $\alpha_{1}=0.24$, $\alpha_{2}=0.30$, $\epsilon=10^{-4}$
  (nonstationary, correlated coexistence). (a) $t=10^{1}$, (b)
  $t=10^{2}$ (c) $t=10^{3}$, (d) $t=10^{4}$. Note that a simulation using
  the parameters of the linear VM, $\alpha_{1}=0.2$, $\alpha_{2}=0.4$,
  would look statistically similar, and also the fluctuations of the
  global frequencies shown in Fig. \ref{x-t-2} are quite similar.}
    \label{posit}
\end{figure}

In order to study the stability of the global dynamics for the different
$\alpha_{1}$, $\alpha_{2}$ settings in the vicinity of the linear voter
point, we have added a small perturbation $\alpha_{0}=\epsilon$.  As the
investigations of Sect. \ref{sec:pertub} have indicated, we should no
longer expect consensus for the perturbated linear VM, but some sort of
coexistence. In fact, we observe an interesting nonstationary pattern
formation we call \emph{correlated coexistence}. Fig. \ref{posit}
(obtained for another range of parameters $\alpha_{1}$, $\alpha_{2}$)
shows an example of this. We find a long-term coexistence of both
species, which is accompanied by a spatial structure formation.  Here,
the spatial pattern remains always nonstationary and the global frequency
randomly fluctuates around a mean value of $x\=0.5$, as shown by
\pic{x-t-2}.
\begin{figure}[htbp]
\centerline{\includegraphics[width=6.5cm]{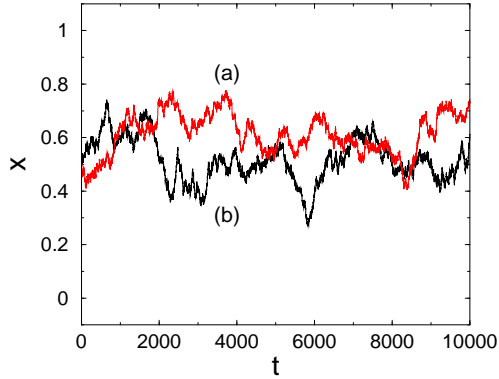}}
    \caption{Global frequency of species 1 vs. time 
      (a) for the linear voter model, $\alpha_{1}=0.2$, $\alpha_{2}=0.4$,
      $\epsilon=10^{-4}$, and (b) for the same setup and parameters as in
      \pic{posit}.  The initial frequency is $x(t$=0)=0.5 for both runs.
    \label{x-t-2}}
\end{figure}

In more specific terms, the regime defined as 'correlated coexistence' is
a paramagnetic phase with finite domain length, typical of partially
phase-separated systems.  We mention that such a regime was also observed
in some related investigations of the VM and other nonlinear spin models
with Ising behavior \citep{dornic:01, deoliveira1993nes}. Also, a similar
transition was observed in the Abrams-Strogatz model
\citep{abrams2003mdl}, where the transition rate is a power $a$ of the
local field. The stability of the solutions then changes at $a=1$, from
coexistence for $a<1$ to dominance for $a>1$.

In order to find out about the range of parameters in the nonlinear VM
resulting in the quite interesting phenomenon of correlated coexistence,
we have varied the parameters $\alpha_{1}$, $\alpha_{2}$ within the
region of positive frequency dependence. Fig. \ref{posit} actually shows
results from a set picked from region (f) in Fig. \ref{phase}, where the
mean-field analysis predicts an asymmetric coexistence of both
species. Obviously, the nonstationarity results from the perturbation
$\epsilon$.

\begin{figure}[htbp]
\centerline{
\fbox{\includegraphics[width=3.5cm]{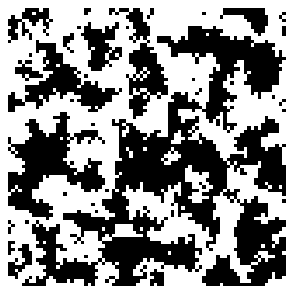}}(a)\hspace{2cm}(b) 
\fbox{\includegraphics[width=3.5cm]{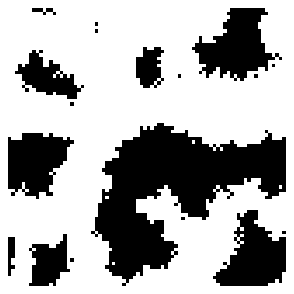}}}
\bigskip
\centerline{
\fbox{\includegraphics[width=3.5cm]{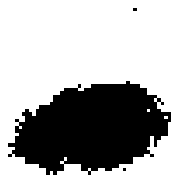}}(c)\hspace{2cm}(d) 
\fbox{\includegraphics[width=3.5cm]{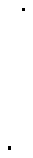}}}
\caption{Spatial snapshots of a positive frequency dependent invasion
  process with $\alpha_{1}=0.1$, $\alpha_{2}=0.3$, $\epsilon=10^{-4}$
  (complete invasion). (a) $t=10^{1}$, (b) $t=10^{2}$ (c) $t=10^{3}$, (d)
  $t=10^{4}$. }
    \label{stoch}
\end{figure}

However, for other sets $\alpha_{1}$, $\alpha_{2}$ in the positive
frequency dependence region the perturbation does \emph{not prevent} the
system from reaching a global ordered state, i.e. invasion of one species
as Fig. \ref{stoch} verifies.
This process is accompanied by a clustering process and eventually a
segregation of both species indicated by the formation of spatial
domains. 
\pic{x-t} depicts the evolution of the global frequency $x(t)$ of species
1 for different intitial frequencies $x(t\=0)$.  In every case, one
species becomes extinct. For $x(t\=0)>0.5$ species 1 is the most likely
survivor, while for $x(t\=0)<0.5$ it is most likely to become
extinct. For $x(t\=0)=0.5$, random events during the early stage
decide about its outcome.

\begin{figure}[htbp]
\centerline{
\fbox{\includegraphics[width=3.5cm,angle=0]{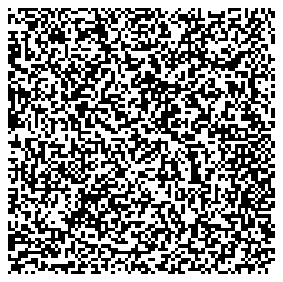}}
}
\caption{Spatial snapshots of a positive frequency dependent invasion
  process with $\alpha_{1}=0.3$, $\alpha_{2}=0.4$, $\epsilon=10^{-4}$
  (random coexistence). The snapshot shown at
  $t=10^{5}$ is statistically equivalent to the initial random state.}
    \label{random}
\end{figure}

On the other hand, the perturbation also does \emph{not induce} an ordered
state as the random coexistence in Fig. \ref{random} shows, which was
again obtained from parameter settings in the region of positive
frequency dependence. So, we conclude that computer simulations of
\emph{positive} frequency dependent processes show three different
dynamic regimes (dependent on the parameters $\alpha_{1}$, $\alpha_{2}$):
(i) complete invasion, (ii) random coexistence, and (iii) correlated
coexistence. While for (i) and (ii) the outcome is in line with the
mean-field prediction shown in Fig. \ref{a1a2}, this does not immediately
follows for (iii). So, we are left with the question whether the
interesting phenomenon of correlated coexistence is just \emph{because}
of the perturbation of some ordered state, or whether it may also exist
\emph{inspite} of $\epsilon$.

\begin{figure}[htbp]
\centerline{\includegraphics[width=6.5cm]{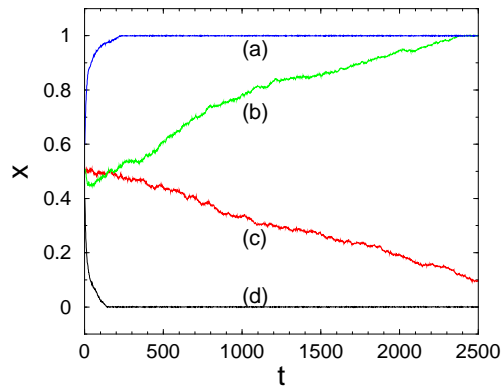}}
\caption{Global frequency of species 1 vs. time for the same setup and
  parameters as in \pic{stoch}. The initial frequencies $x(t$=0) of the
  four different runs are: (a) 0.6, (b) 0.5, (c) 0.5, (d) 0.4.}.
    \label{x-t}
\end{figure}

We just add that for \emph{negative} frequency dependent invasion,
\eqn{eps-alpha}, the the spatial pattern remains random, similar to
\pic{random}. Furthermore, regardless of the initial frequency $x(t\=0)$
, on a very short time scales, a global frequency $x^{\mathrm{stat}}$=0.5
is always reached. That means we always find \emph{coexistence} between
both species. We conclude that for \emph{negative} frequency dependence
$x^{\mathrm{stat}}$=0.5 is the only stable value, which is in agreement
with the mean-field prediction, whereas for \emph{positive} frequency
dependence the situation is not as clear.

\section{Derivation of a phase diagram}
\label{4}

To answer the question what ranges of $\alpha_{1}$, $\alpha_{2}$
eventually lead to what kind of macroscopic dynamics, we now make use of
the pair approximation already derived in Sect. \ref{3.2} as a first
correction to the mean-field limit. Here, we follow a two-step strategy:
First, we investigate how well the pair approximation, eqs.
(\ref{eq:master2d}), (\ref{eq:doublet2d}), (\ref{eq:master2dc}) of the
macroscopic dynamics, \eqn{x-fin}, predict the global quantities
$\mean{x}$ and $c_{1|1}$.  In order to specify the network topology, we
use again the CA described above.  Second, we use the pair approximation
to derive a phase diagram in the case of local interaction. Eventually,
we test whether these findings remain stable against perturbations of the
ordered state.  All predictions are tested by comparison with computer
simulations of the microscopic model, from which we calculate the
quantities of interest and average them over 50 runs.

\subsection{Global frequencies and spatial correlations}
\label{4.1}

Here we have to distinguish between the three different dynamic regimes
already indicated in Sect. \ref{2.4}.

Regime (i), \emph{complete invasion}, is characterized by fixed points of
the macroscopic dynamics of either $\{\mean{x}, c_{1|1}\}=\{1,1\}$ or
$\{\mean{x}, c_{1|1}\}=\{0,0\}$. The CA simulations as well as also the
pair approximation of the dynamics quickly converge to one of these
attractors, dependent on the initial conditions.

Regime (ii), \emph{random coexistence}, has only one fixed point,
$\{\mean{x}, c_{1|1}\}=\{0.5,0.5\}$, to which the CA simulations quickly
converge. The pair approximation converges to $\mean{x}=0.5$ after some
initial deviations from the CA simulation, i.e. it relaxes on a different
time scale ($t>40$), but is correct in the long run.  The approximation
of the local correlation $c_{1|1}$ shows some deviations from the
predicted value, $c_{1|1}=0.5$. We have tested the case of random
coexistence for various parameter values and found values for $c_{1|1}$
between 0.4 and 0.6. The discrepancy is understandable, since in the case
of long-term coexistence some of the spatial patterns flip between two
different random configurations with high frequency.  Thus, while the
global frequency settles down to 0.5, the microscopic dynamics is still
nonstationary.

\begin{figure}[htbp]
{\includegraphics[width=6.5cm,clip]{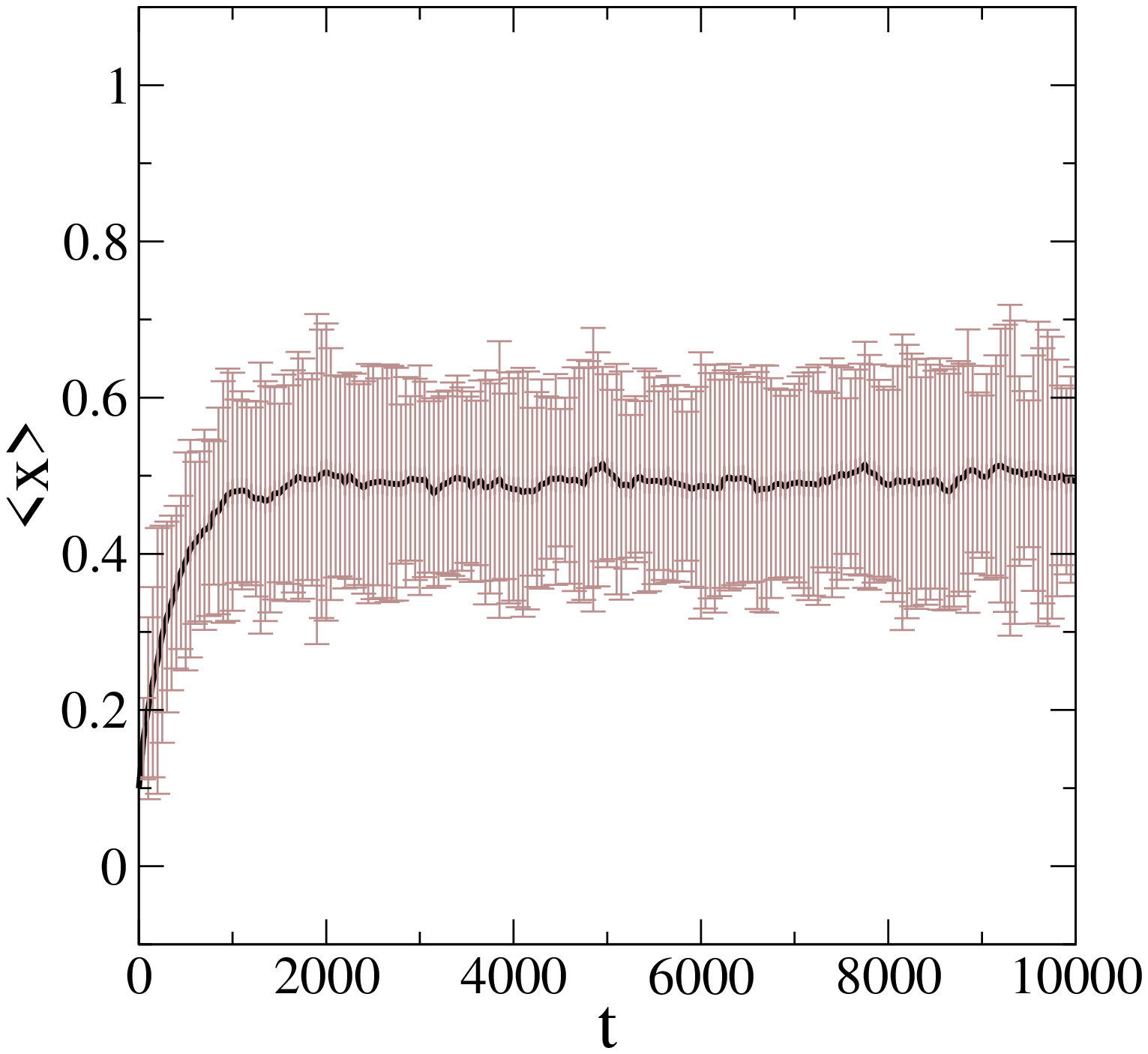}}\hfill
{\includegraphics[width=6.5cm]{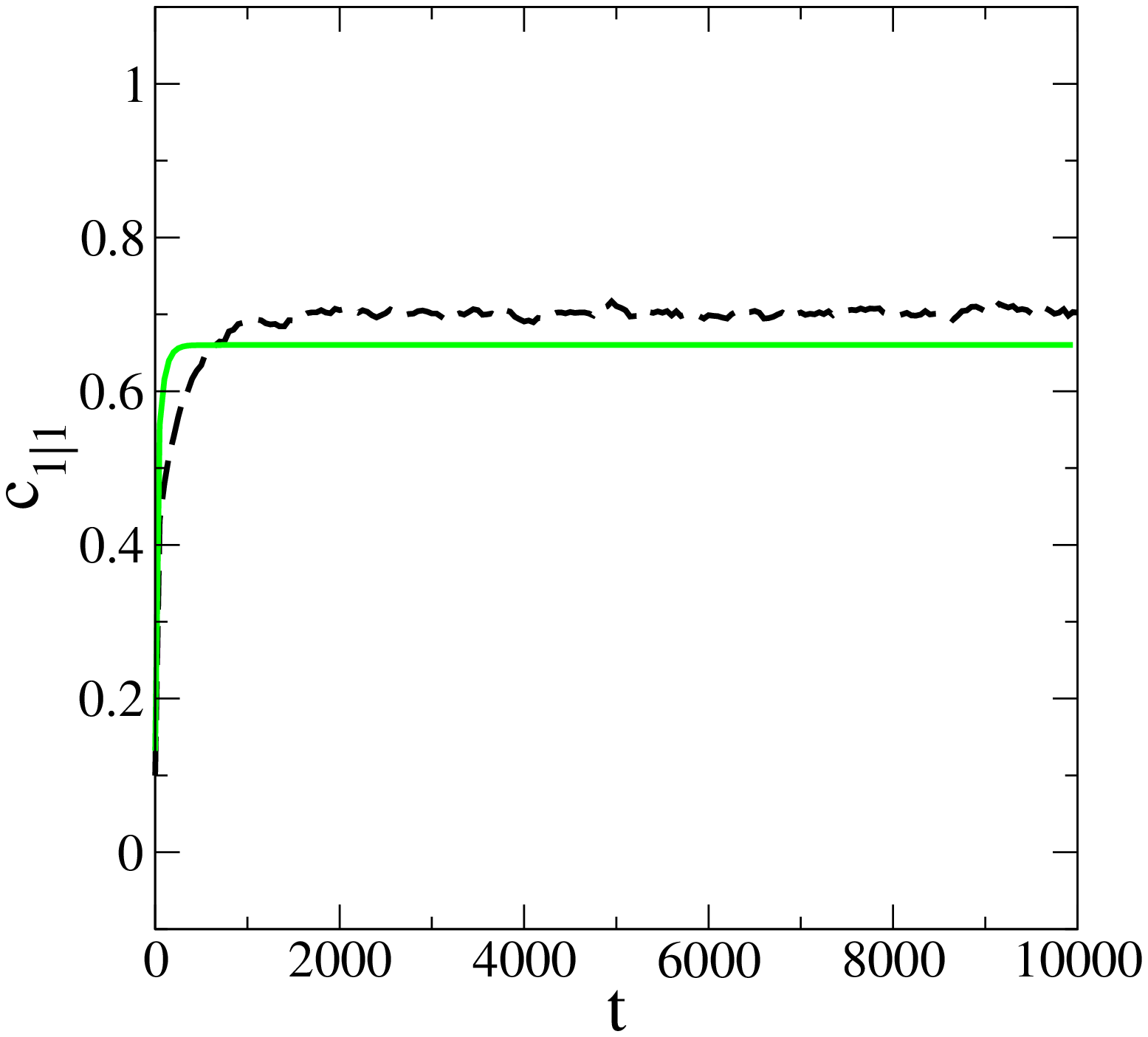}}
  \caption{
Global frequency $\mean{x}$ with min-max deviations (top) and
    spatial correlation $c_{1|1}$ (bottom)  for the case of correlated
    coexistence. The two curves shown in the bottom part result from averaging over 50 CA
    simulations (black dotted line) and from the pair
approximation (green solid line) 
 The parameters
    are as in 
     Figs. \ref{posit}, \ref{x-t-2}.
     (top) 
     \label{variat}}
 \end{figure}

Regime (iii), \emph{correlated coexistence}, the most interesting one, is
chacterized by an average global frequency of $\mean{x}=0.5$ again,
however the existing local correlations lead to a much higher value of
$c_{1|1}>0.6$. This is shown in Fig. \ref{variat}, where we find
$c_{1|1}\approx 0.7$ from the CA simulations and $c_{1|1}\approx 0.65$
from the pair approximation. I.e., for the case of spatial domain
formation the long-range correlations can be well captured by the pair
approximation, whereas this was less satisfactory for the short-range
correlations of the random patterns.

\subsection{Determining the phase boundary}
\label{4.3}
\begin{figure}[htbp]
{\includegraphics[width=6.5cm]{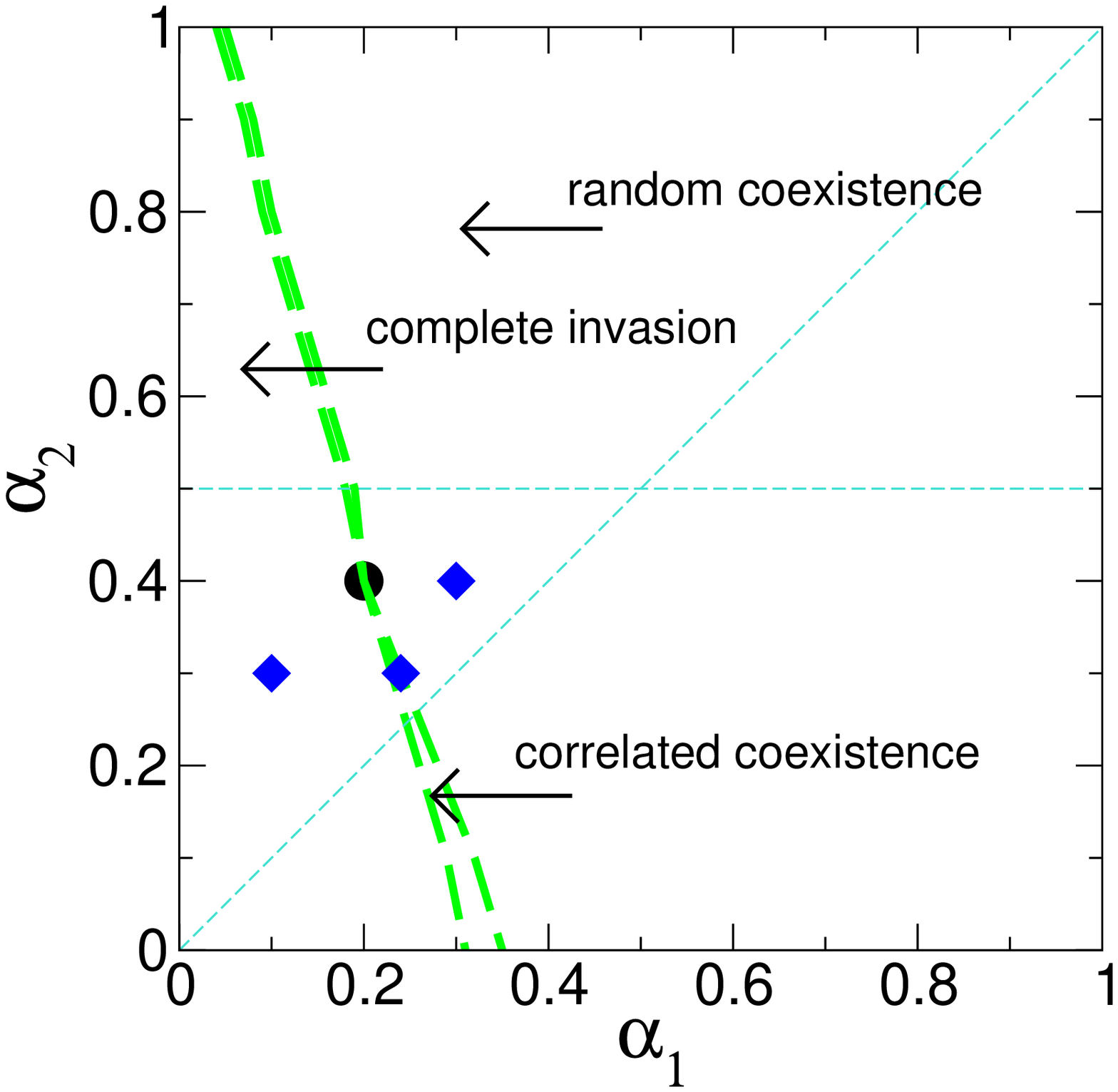}} \hfill
{\includegraphics[width=6.5cm]{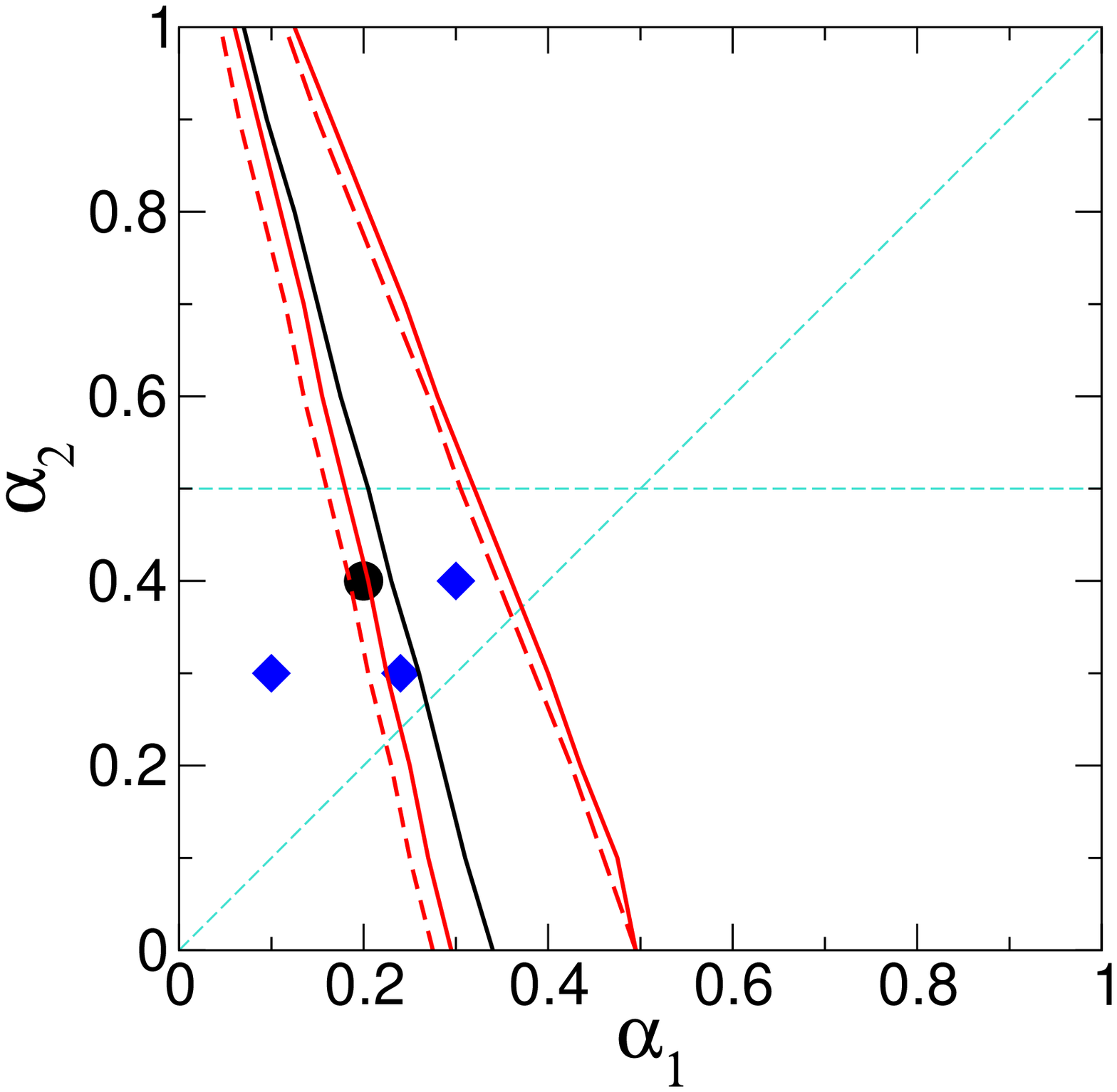}} 
\caption{Phase diagram of the nonlinear voter model: (top) CA
  simulations, averaged over 50 runs ($c_{1|1}=0.7$), (bottom) pair
  approximation. Phase boundaries for $\epsilon=0$ resulting from $c_{1|1}=0.65$ (upper
  limit): left -- red solid line, right -- black solid line, for
  comparison phase boundaries resulting from  $c_{1|1}=0.60$ (lower
  limit): left -- red solid line (identical with $c_{1|1}=0.65$), right
  -- red solid line on the far right. Dashed red lines indicate the shift
  of the phase boundaries for $c_{1|1}=0.60$ if  $\epsilon=0.02$. 
  The linear voter point (0.2,0.4) is indicated by $\bullet$.  Further,
  $\Diamond$ marks those three parameter sets from the positive frequency
  dependence region where CA simulations are shown in Figs.  \ref{stoch},
  \ref{posit}, \ref{random} (see also global frequency $\mean{x}$ and
  local correlation $c_{1|1}$ in Figs.  \ref{x-t-2}, \ref{x-t},
  \ref{variat}). The straight dashed lines mark the different parameter
  areas given in \eqn{eps-alpha} which are also shown in
  Fig. \ref{region}.}
\label{phasepl}
\end{figure}

The insights into the dynamics of the nonlinear VM derived in this paper
are now summarized in a phase diagram that identifies the different
parameter regions $(\alpha_{1},\alpha_{2})$ for the possible dynamic
regimes identified in the previous sections.  In order to find the
boundaries between these different regimes, we carried out CA simulations
of the complete parameter space, $0\leq (\alpha_{1},\alpha_{2})\leq
1$. Precisely, for every single run the long-term stationary values of
$x$ and $c_{1|1}$ were obtained and then averaged over 50 simulations. As
described above, the three different regimes could be clearly separated
by their $\{\mean{x},c_{1|1}\}$ values, which were used to identify the
phase boundary between the different regimes.  The outcome of the CA
simulations is shown in the phase diagram of Fig. \ref{phasepl} (top) and
should be compared with Fig. \ref{phase}, which results from the
mean-field analysis in Sect. \ref{3.2} and thus neglects any kind of
correlation.

Instead of the six regions distinguished in Fig.  \ref{phase}, in the
local case we can distinguish \emph{two} different regions divided by one
separatrix: The parameter region left of the separatrix refers to the
\emph{complete invasion} of one of the species, with \emph{high} local
correlations during the evolution.  In the CA simulations, we observe the
formation of domains that grow in the course of time until exclusive
domination prevails. Asymptotically, a \emph{stationary} pattern is
observed, with $\mean{x}=1$ (or 0) and $c_{1|1}=1$ (or 0) (cf the
simulation results show in Figs.  \ref{stoch}, \ref{x-t}).  I.e., the
system converges into a 'frozen' state with no dynamics at all.  The
region to the right of the separatrix refers to \emph{random coexistence}
of both species with $\mean{x}=0.5$ and \emph{no} local correlations,
i.e. $c_{1|1}=0.5$.  In the CA simulations, we observe
\emph{nonstationary} random patterns that change with high frequency
(cf. Fig. \ref{random}).

Both regions are divided by a \emph{separatrix}.  As shown in
\pic{phasepl} (top), the separatrix is divided into two pieces by the
linear voter point, (0.2,0.4).  Above that point, the separatrix is very
narrow, but below the voter point it has in fact a certain extension in
the parameter space. Looking at the dynamics \emph{on} the separatrix, we
find that the mean frequency is $\mean{x}=0.5$ both above and below the
voter point (see also Fig.  \ref{variat}, top).  The local correlation
$c_{1|1}=0.7$ holds within the whole extended area of the separatrix
which identifies it as the region of \emph{correlated coexistence} (see
also Fig.  \ref{variat}, bottom).

The question is how well this phase diagram can be predicted by using the
pair approximation of the dynamics, described by
eqs. (\ref{eq:master2d}), (\ref{eq:master2dc}). For a comparison, the
coupled equations were numerically solved to get the asymptotic solutions
for $\{\mean{x},c_{1|1}\}$ (which looks complicated but is numerically
very fast and efficient). In order to distinguish between the three
regimes, we have to define a critical $c_{1|1}$ value for the
correlations. Whereas in the CA simulations $c_{1|1}=0.7$ indicated a
correlated nonstationary coexistence, this value was never reached using
the pair approximation (see Fig. \ref{variat}, bottom), so $c_{1|1}\leq
0.65$ could be regarded as an \emph{upper limit} for the case of
correlated coexistence.  Random coexistence, on the other end, yielded
$c_{1|1}=0.5$ in the CA simulations and a value between 0.4 and 0.6 in
the pair approximation. So, given suitable initial conditions, we can
regard $c_{1|1}\geq 0.6$ as a \emph{lower limit}, to identify
correlated coexistence. The results are shown in Fig. \ref{phasepl}
(bottom) which shall be compared with the phase diagram above (Fig.
\ref{phasepl}, top).

Fig. \ref{phasepl} (bottom) shows the influence of the $c_{1|1}$
threshold value. With the lower limit, we find a quite broad region of
correlated coexistence, which for example also includes the point
$(0.3,0.4)$ for which a random coexistence in the CA simulations was
shown in Fig. \ref{random}. Using the upper limit $c_{1|1}=0.65$
results in a much smaller region of correlated coexistence. Comparing
this with the CA simulations above, we can verify that the pair
approximation correctly predicts the extended region below the voter
point and also shows how it becomes more narrow above the voter point.
One should note that the left border of the separatrics is not affected
by the threshold value, whereas the right border shifts considerably.
The left border also contains the voter point (independent of the
$c_{1|1}$ threshold value), for which a complete invasion can be
observed.

Therefore, it is quite interesting to look into changes of the phase
diagram if additional perturbations are considered (see also
Sect. \ref{sec:pertub}). Fig. \ref{phasepl} (bottom) shows (for the
threshold $c_{1|1}=0.6$) that this does \emph{not} affects the existence
of the three dynamic regimes and most notably of the extended separatrix
below the voter point, but only shifts the boundaries toward the left,
dependent on the value of $\alpha_{0}=\epsilon$ (this can be also
verified for $c_{1|1}=0.65$ but is omitted here, to keep the figure
readable). Thus, the consideration of perturbations in the phase diagram
reveals that it is indeed the \emph{nonlinearity} in the voter model
which allows for the interesting phenomenon of the correlated coexistence
and \emph{not} just the perturbation.

A closer look into Fig.  \ref{phasepl} (bottom) also shows that in the
perturbated phase diagram the voter point no longer lies on the boundary
towards the region of complete invasion but clearly \emph{within} the
region of correlated coexististence. This is in agreement with the
findings in Sect. \ref{sec:pertub} which showed that for the linear VM
complete invasion is an unstable phenomenon and changes into correlated
coexistence for finite $\epsilon$.

\section{Discussion and Conclusions}
\label{5}

In this paper, we investigated a local model of frequency dependent
processes, which for example models the dynamics of two species $\{0,1\}$
in a spatial environment. Individuals of these species (also called
'voters') are seen as nodes of a network assumed as homogeneous in this
paper (i.e. all nodes have the same number of neighbors, $m$). The basic
assumption for the microscopic dynamics is that the probability to occupy
a given node with either species $0$ or $1$ depends on the frequency of
this species in the immediate neighborhood. Different from other
investigations, we have counted in the state of the center node as well
(see Sect. \ref{sec:vm}) and have further considered a \emph{nonlinear
  response} of the voters to the local frequencies.

Studies of a nonlinear version of the traditional voter model (without
counting the state of the central node and with sequential dynamics) have
already been analyzed before. Ref. \citep{molofsky99}, as pointed out
before, is closest to our investigations, but restricted itself to the
mean-field analysis and computer simulations of the 2d case, to obtain a
phase diagram similar to Fig. \ref{phasepl}
(top). \citep{muehlenb-hoens-02}, on the other hand, have provided a
Markov analysis which is restricted only to very small CA.  The
two-parameter model in \citep{deoliveira1993nes} is for $y=1$ a nonlinear
voter model that exhibits at the voter point ($x=1/2$) a transition from
a ferromagnetic phase, i.e. invasion, for $x>1/2$ to a paramagnetic phase
(correlated coexistence) for $x<1/2$.  Also the case of the 'perturbated
linear voter model' is included in the model, for $x=1/2$ and $y<1$.
Similar results are also presented in \citep{drouffe1999poa}, which
points out relations to random branching processes, and in
\citep{castello2006odt}, where the emphasis is on investigations of the
interface density, to describe the coarsening process. A recent paper
\citep{vazquez08-unp} also shows for spin systems with two symmetric
absorbing states (such as the VM) that the macroscopic dynamics only
depends on the first derivatives of the spin-flip probabilities.

In our paper, we set out for a formal approach that allows to derive the
dynamics on different levels: (i) a \emph{stochastic dynamics} on the
microlevel, which is used for reference computer simulations but also
allows a derivation of the (ii) \emph{macroscopic dynamics} for the key
variables, given in terms of differential equations. This macroscopic
dynamics is then analysed by two different approximations, (i) a
\emph{mean-field approximation} that neglects any local interaction in
the network, and (ii) a \emph{local approximation} considering
correlations between pairs of nearest neighbors. In order to test the
validity of these approximations, we compare their preditions with the
averaged outcome of the microscopic computer simulations. We like to
emphasize that our approach is general enough to be applied to various
forms of \emph{frequency dependent processes} on homogeneous networks
with different number of neighbors. Even if a two-dimensional regular
lattice is used to illustrate the dynamics, the approach is not
restricted to that.

Our main result, in addition to the general framework for nonlinear
frequency dependent processes, is the derivation of a \emph{phase
  diagram} using the \emph{pair approximation} derived in this paper.
This approach predicts correctly both the type of the dynamics and the
asymptotic values of the key variables dependent on the possible
nonlinearities for the case of \emph{local interaction}, $m=4$. The
predicted phase diagram was verified by comparision with extensive
microscopic computer simulations rastering the whole parameter
space. While the structure of the phase diagram was already known from
previous computer simulations presented in \citep{molofsky99} we could
demonstrate that the pair approximation works very well both for
predicting the correct phase boundaries and the dynamics within these
phases. It should be noticed that the pair approximation is a valuable
tool, particularly with respect to computational efforts.  The computer
simulations are much more timeconsuming, since the results of the
different runs have to be averaged afterwards.  The pair approximation,
on the other hand, is based on only 2 coupled equations and therefore
needs less computational effort. In the following, we discuss some of the
interesting findings.

\emph{The region of correlated coexistence:} Analysing the nonlinear VM
with \emph{local interaction} has shown that there are in fact only
\emph{three} different dynamic regimes dependent on the nonlinearities
$(\alpha_{1},\alpha_{2})$: (i) complete invasion, (ii) random
coexistence, and (iii) correlated coexistence. The first one is already
known as the standard behavior of the \emph{linear} VM. Consequently the
only interesting feature, namely the time to reach the ordered state
dependent on the network size and topology, has been the subject of many
investigations \citep{castellano2005, maxi2, suchecki2005}. Number (ii),
on the other hand, only leads to trivial results as no real dynamics is
observed. Thus, the most interesting regime is (iii) correlated
coexistence, which can be found in a small, but not negligible parameter
region below the voter point. This region separates the two dominant
regimes (i) and (ii) and therefore was called a separatrix here. Going
over from the right to the left side of the phase diagram in that region,
we notice a transition from 0.5 to 1.0 (or 0.0 respectively) in the mean
frequency, and from 0.5 to 0.7 to 1.0 in the local correlations.  Thus,
in fact $c_{1|1}$ separates the two dynamic regimes (i) and (ii) (below
the voter point).  For parameters chosen from that region, we find in the
CA simulations a long-term and nonstationary \emph{coexistence} between
the two species as on the \emph{right} side of the phase diagram. But we
also find the long-range spatial correlations that lead to the formation
of spatial domains as shown e.g. in \pic{posit} -- which is
characteristic for the \emph{left} side of the phase diagram.  The
spatial pattern formation is also indicated by large fluctuations of
$\mean{x}$ shown in \pic{variat}(top).
A single run, as shown in \pic{x-t-2}, clearly indicates the long-term
nonstationary coexistence of both species. 

We emphasize that the separatrix between the two dynamic regimes is well
predicted by the macroscopic dynamics resulting from the pair
approximation (as can be clearly seen in \pic{phasepl}).  Most
importantly, we could verify that the correlated coexistence of both
species is not simply the effect of an additional perturbation, but
results from the nonlinear interaction.

\emph{Comparison with the mean-field phase diagram:} In our paper, the
mean-field approximation plays the role of a reference state used to
demonstrate the differences of the local analysis. The phase diagram of
Fig. \ref{phase} distinguishes between six different regions, whereas the
one in the local case, Fig. \ref{phasepl} (top) shows only three.
Comparing the two phase diagrams, we realize that the most interesting
regions in \pic{phase}, namely (c) and (f), have simply collapsed into
the separatrix shown in Fig.  \ref{phasepl}.  The region (c) of unstable
asymmetric coexistence or multiple outcome, respectively (see
Sect. \ref{3}), relates to the separatrix line above the voter point. It
should be noticed that the phase diagram for local interaction,
Fig. \ref{phasepl}, correctly predicts that the deterministic behavior
for $(\alpha_{1},\alpha_{2})=(0,1)$ leads to complete invasion (see
Sect. \ref{4.3} and Fig. \ref{det01}). 

The region (f) of stable asymmetric coexistence relates to the extended
area of the separatrix shown below the voter point in \pic{phasepl},
where we still see a coexistence of both species - but the asymmetry
between the two species relates to their changing dominance over time, as
Figs. \ref{x-t-2}, \ref{variat} (top) clearly illustrate.  We conclude
that in the local case no regions of stationary \emph{and} asymmetric
coexistence between the two species exist, as was predicted by the
mean-field analysis. However, we find a (small but extended) region
\emph{on} the separatrix that shows the \emph{nonstationary} and
asymmetric coexistence of the two species for \emph{single} realizations
(which results in a symmetric coexistence averaged over runs,
$\mean{x}=0.5$, see Fig. \ref{variat}, top).

\emph{The role of positive frequency dependence:} The possible nonlinear
responses in frequency dependent processes can be distinguished in four
parameter areas of positive and negative frequency dependence and
positive and negative allee effects, as Fig. \ref{region} shows. Previous
investigations \citep{molofsky99} assigned a dynamic leading to invasion
to a positive frequency dependence, while associating a spatial
coexistence with negative frequency dependence. Our investigations have
shown that such an assignment does not unambiguously hold.  In
particular, a random coexistence can be found for \emph{negative}
frequency dependent dynamics as well as for \emph{positive} frequency
depencence, which was so far assigned to complete invasion only
\citep{molofsky99}. On the other hand, complete invasion is not observed
only for positive frequency dependence, but also for positive and
negative allee effects. A random spatial coexistence can be found for
positive and negative allee dynamics as well. The only case where just
one dynamic regime can be observed is the case of negative frequency
dependence.  We note, however, that the nonstationary long-term
coexistence with spatial pattern formation occurs both for the positive
frequency dependence and the negative allee dynamics, given that the
parameters are chosen from the most interesting zone of the separatrix
below the voter point.

In conclusion, the region of positive frequency dependence bears in fact
a much more richer dynamics, as it is transected by the separatrix we
identified in the local analysis and thus shows all three types of
dynamics we could identify for nonlinear voter models, namely (i)
complete invasion of one of the species via the formation of large
domains, (ii) long-term coexistence of both species with random
distribution, (iii) long-term coexistence of both species with formation
of nonstationary domains. However, the most interesting regime (iii) is
\emph{not} restricted to positive frequency dependent processes, but can
be also found for some negative allee effects. 

We summarize our findings by pointing out that \emph{nonlinear} VM show
indeed a very rich dynamics which was not much investigated yet. In
addition to the phenomenon of complete invasion (or consensus) which
occurs also beyond the linear VM, we find most interesting that certain
parameter settings lead to a dynamics with nonstationarity and long-term
correlations.  Thinking about possible applications of the VM, we see
that in particular this region has the potential to model relevant
observations, be it the temporal dominance of certain species in a
habitat or the temporal prevalence of certain opions (or political
parties) in a social system. The nonstationarity observed gives rise to
the prediction that such dominance may not be the end, and change happens
(even without additional perturbation).

\subsection*{Acknowledgments}
The authors want to thank Thilo Mahnig and Heinz M\"uhlenbein for
discussions on an early version of this paper.

\appendix
\section*{Appendix}
\label{A}

Here we derive some explicit expressions for the three equations of the
pair approximation discussed in Sect. \ref{3.2}, for the global frequency
$\mean{x}$ (\eqn{macro-pair}), the doublet frequency $\mean{x_{1,1}}$ and
the correlation term $c_{1|1}$, \eqn{eq:local1}). The equations are
derived for the neighborhood $m=4$. We use the notation $x\equiv
\mean{x}$. Using \eqn{eq:cond_prob} and the transition rates of
\eqn{trans2}, we find for $\mean{x}$, \eqn{macro-pair} in pair
approximation:
\begin{equation}
\label{eq:master2d}
\begin{aligned}
  \frac{dx}{dt}= & \epsilon \Big[ \frac{1}{(1-x)^3}(1-2x+xc_{1|1})^4 - x
  {c_{1|1}}^4 \Big] \\
& +4\alpha_1 \left[ \frac{x}{(1-x)^3} (1-2x+xc_{1|1})^3 (1-c_{1|1})
  \right. \\
& \qquad \quad - x (1-c_{1|1}) {c_{1|1}}^3 \Big] \\
& + 6 \alpha_2 \left[ \frac{x^2}{(1-x)^3}
(1-2x+xc_{1|1})^2 (1-c_{1|1})^2 \right. \\ 
& \qquad \quad - x (1-c_{1|1})^2 {c_{1|1}}^2 \Big] \\
& +(1-\alpha_1) \left[
\frac{x^4}{(1-x)^3} (1-c_{1|1})^4 - x (1-c_{1|1})^4 \right] \\
& +4(1-\alpha_2)\left[ \frac{x^3}{(1-x)^3}(1-2x+xc_{1|1}) (1-c_{1|1})^3
  \right. \\
& \qquad \quad - x   (1-c_{1|1})^3 c_{1|1} \Big]
\end{aligned}
\end{equation}
We note that $c_{1|1}=c_{1|0}=x$ and $c_{0|0}=c_{0|1}=1-x$ in the mean-field
limit, in which case \eqn{eq:master2d} reduces to \eqn{eq:mean_field}.

In order to calculate the time derivative of the doublet frequency $\mean{
  x_{1,1} }$ we have to consider how it is affected by changes of $\sigma$ in
a specific occupation pattern of size $m=4$,
$\ul{\sigma}^{0}=\{\sigma,\sigma_{1},\sigma_{2},\sigma_{3},\sigma_{4}\}$,
considering the $\sigma_{j}$ as constant.  Again, in a frequency dependent
process it is assumed that the transition does not depend on the exact
distribution of the $\sigma_{j}$, but only on the frequency of a particular
state $\sigma$ in the neighborhood.  Let $S_{\sigma,q}$ describe a
neighborhood where the center node in state $\sigma$ is surrounded by $q$
nodes of the same state $\sigma$.  For any given $q\leq m$, there are
${m}\choose q$ such occupation patterns. The global frequency of neighborhood
$S_{\sigma,q}$ is denoted as $x_{\sigma,q}$ with the expectation value
$\mean{x_{\sigma,q}}$. Obviously, $x_{\sigma,q}$ can be calculated from the
global frequencies $x_{\sigma,\ul{\sigma}^{\prime}}$ of all possible
occupation distributions $\ul{\sigma}^{\prime}$ ( \eqn{sigm01}), that match
the condition
\begin{equation}
  \label{cond}
  z^{\sigma}=\sum_{j=1}^{m} \delta_{\sigma,\sigma_{j}} \;:=\;q
\end{equation}
i.e. it is defined as
\begin{equation}
  \label{s-m}
  x_{\sigma,q}= \sum_{\ul{\sigma}^{\prime},z^{\sigma^{\prime}}\=q}
  x_{\sigma,\ul{\sigma^{\prime}}}
\end{equation}
Regarding the possible transitions, we are only interested in changes of the
doublet (1,1), i.e. transitions $(1,1)\to(0,1)$ or $(0,1)\to(1,1)$.  The
transition rates shall be denoted as $w\big((0,1)|(1,1),S_{\sigma,q}\big)$ and
$w\big((1,1)|(0,1),S_{\sigma,q}\big)$ respectively, which of course depend on
the local neighborhood $S_{\sigma,q}$. With this, the dynamics of the expected
doublet frequency can be described by the rate equation:
\begin{equation}
\label{d_frequency2}
\begin{aligned}
\frac{d}{dt}\mean{x_{1,1}}(t)= \sum_{q=0}^{m} \Big [&
w\big((1,1)|(0,1),S_{0,q}\big)\mean{x_{0,q}} \\ & -
w\big((0,1)|(1,1),S_{1,q}\big)\mean{x_{1,q}} \Big]
\end{aligned}
\end{equation}
In order to specify the transition rates of the doublets
$w\big((\sigma^{\prime},1)|(\sigma,1),S_{\sigma,q}\big)$, with
$\sigma^{\prime}=1-\sigma$ and $\sigma\in \{0,1\}$, we note that there are
only 10 distinct configurations of the neighborhood. Let us take the example
$\ul{\sigma}^{0}=\{1,1,1,1,1\}$.  A transition $1\to 0$ of the center node
would lead to the extinction of 4 doublets $(\sigma,\sigma_{j})=(1,1)$. On the
other hand, the transition rate of the center node is $\epsilon$ as known from
\eqn{trans2}.  This would result in $w\big((0,1)|(1,1),S_{1,4}\big)\propto
4\epsilon$. However, for a lattice of size $N$ the number of doublets is $2N$,
whereas there are exactly $N$ neighborhoods $\ul{\sigma}^{0}$. Therefore, if
we apply the transition rates of the single nodes, \eqn{trans2}, to the
transition of the doublets, their rates have to be scaled by 2. Similarly, if
we take the example $\ul{\sigma}^{0}=\{0,1,1,1,0\}$, a transition of the
center node $0\to1$ would occur at the rate $1-\alpha_{2}$ and would create 3
new doublets. Applying the scaling factor of 2, we verify that
$w\big((1,1)|(0,1),S_{0,1}\big)=3/2 \;(1-\alpha_{2})$. This way we can
determine the other possible transition rates:
\begin{equation}
\label{eq:tp_2dn}
\begin{aligned}
w\big((0,1)|(1,1),S_{1,4}\big)& =2\epsilon \\ 
w\big((0,1)|(1,1),S_{1,3}\big)&=\frac{3}{2}\alpha_1 \\ 
w\big((0,1)|(1,1),S_{1,2}\big)&=\alpha_2 \\ 
w\big((0,1)|(1,1),S_{1,1}\big)&=\frac{1}{2}(1-\alpha_2) \\ 
w\big((0,1)|(1,1),S_{1,0}\big)&=0 \\ 
w\big((1,1)|(0,1),S_{0,4}\big)&=0\\
w\big((1,1)|(0,1),S_{0,3}\big)&=\frac{1}{2}\alpha_1 \\
w\big((1,1)|(0,1),S_{0,2}\big)&=\alpha_2 \\
w\big((1,1)|(0,1),S_{0,1}\big)&=\frac{3}{2}(1-\alpha_2) \\
w\big((1,1)|(0,1),S_{0,0}\big)&=2(1-\alpha_1)
\end{aligned}
\end{equation}
Note that two of the transition rates are zero, because the respective
doublets (1,1) or (0,1) do not exist in the assumed neighborhood.  Finally, we
express the $\mean{x_{\sigma,q}}$ in \eqn{d_frequency2} by the
$\mean{x_{\sigma,\ul{\sigma}^{\prime}}}$ of \eqn{s-m} and apply the pair
approximation, \eqn{eq:pairapprox}, to the latter one. This way, we arrive at
the dynamic equation for $\mean{x_{1,1}}$:
\begin{equation}
\label{eq:doublet2d}
\begin{aligned}
\frac{d\mean{ x_{1,1}}}{dt}=& -2\epsilon x {c_{1|1}}^4 \\
& +2\alpha_1 \left[\frac{x}{(1-x)^3}
(1-x+xc_{1|1})^3 (1-c_{1|1}) \right.\\
& \qquad \quad - 3 x (1-c_{1|1}) {c_{1|1}}^3 \Big] \\
& + 6 \alpha_2 \left[\frac{x^2}{(1-x)^3}
(1-x+xc_{1|1})^2 (1-c_{1|1})^2 c_{1|1}^{2} \right.\\
& \qquad \quad -x (1-c_{1|1})^2 \Big] \\
& + 2(1-\alpha_1) \left[
\frac{x^4}{(1-x)^3} (1-c_{1|1})^4 \right. \\
& \qquad \quad -x (1-c_{1|1})^3 \Big] \\
& +2(1-\alpha_2) \left[
\frac{x^3}{(1-x)^3}3 (1-2x+xc_{1|1})\times  \right. \\
& \qquad \quad  \times (1-c_{1|1})^3 - x (1-c_{1|1})^3
c_{1|1} \Big]
\end{aligned}
\end{equation}
The third equation, \eqn{eq:local1}, for the correlation term $c_{1|1}$ can be
obtained in explicte form by using \eqn{eq:master2d} for $\mean{x}$ and
\eqn{eq:doublet2d} for $\mean{ x_{1,1}}$: 
\begin{equation}
\label{eq:master2dc}
\begin{aligned}
\frac{dc_{1|1}}{dt}=& - \epsilon \left(\frac{c_{1|1}}{x(1-x)^3}
(1-2x+xc_{1|1})^4+ {c_{1|1}}^5 -2 {c_{1|1}}^4 \right)\\
&+\alpha_1(1-c_{1|1})
\Big[ {c_{1|1}}^3(4c_{1|1}-6)  \\ 
& \quad -2\frac{1}{(1-x)^3}
(1-x+xc_{1|1})^3(2c_{1|1}-1)\Big]\\
&+ 6\alpha_2 (1-c_{1|1})^3 \left[\frac{x}{(1-x)^3} (1-x+xc_{1|1})^2
-c_{1|1}\right]\\
& +(1-\alpha_1)(1-c_{1|1})^4 \left[\frac{x^3}{(1-x)^3} (2-c_{1|1})
+c_{1|1}\right] \\
& +(1-\alpha_2)(1-c_{1|1})^3 \Big[2c_{1|1}(2c_{1|1}-1) \\
& \quad +
  \frac{x^2}{(1-x)^3}(1-2x+xc_{1|1})(6-4c_{1|1}) \Big]
\end{aligned}
\end{equation}

\end{document}